\documentclass[letter]{sig-alternate}
\usepackage{xspace}
\usepackage{graphicx}
\usepackage[usenames,dvipsnames]{color}
\usepackage[normalem]{ulem}
\usepackage{subfig}
\usepackage{times}
\newcommand{\subparagraph}{}

\usepackage{ulem}
\newcommand{\xref}[1]{\S\ref{#1}}
\newcommand{\minisec}[1]{\textbf{#1:}}

\newcommand{\match}{a^{2}\textrm{-match-}a^{1}}
\newcommand{\nonmatch}{a^{2}\textrm{-non-match-}a^{1}}
\newcommand{\matchs}{\textrm{match}}
\newcommand{\nonmatchs}{\textrm{non-match}}
\newcommand{\impers}{a^{2}\textrm{-impersonate-}a^{1}}
\newcommand{\nonimpers}{a^{2}\textrm{-non-imperso-}a^{1}}

\newcommand{\change}[1]{{\emph{\color{red}#1}}}

\newcommand{\comment}[1]{{\color{red}\textit{(#1)}}}
\newcommand{\oana}[1]{{\emph{\color{red}#1 --Oana }}}

\newcommand{\update}[1     ]{{\color{MidnightBlue} #1}}

\newcommand{\toremove}[1]{{\sout{#1}}}

\newcommand{\nb}{850\xspace}
\newcommand{\smallscale}{\textsc{Random-Sampled}\xspace}
\newcommand{\largescale}{\textsc{Emulated-Large}\xspace}

\newcommand{\linkernb}{\textsc{Linker-NB}\xspace}
\newcommand{\linkersvm}{\textsc{Linker-SVM}\xspace}
\newcommand{\hybridsvm}{\textsc{Linker-SVM+}\xspace}

\newcommand{\profile}{profile\xspace}
\newcommand{\profiles}{profiles\xspace}
\newcommand{\matching}{matching\xspace}
\newcommand{\nonmatching}{non-matching\xspace}
\newcommand{\nonmatchingprofile}{non-matching profile\xspace}
\newcommand{\nonmatchingprofiles}{non-matching profiles\xspace}
\newcommand{\matchingprofile}{matching profile\xspace}
\newcommand{\matchingprofiles}{matching profiles\xspace}
\newcommand{\tname}{\textsc{ACID}\xspace}
\newcommand{\sysname}{\textsc{Linker}\xspace}
\newcommand{\filter}{\textsc{Filter}\xspace}
\newcommand{\disambiguator}{\textsc{Disambiguator}\xspace}
\newcommand{\confidence}{\textsc{Guard}\xspace} 
\newcommand{\psysname}{\textsc{topmatch}\xspace}
\newcommand{\sysnameS}{\textsc{Scalable-Linker}\xspace}
\newcommand{\sysnameD}{\textsc{Disambiguator}\xspace}
\newcommand{\datasetFF}{\textsc{Dataset~FF}\xspace}
\newcommand{\datasetG}{\textsc{Dataset~G+}\xspace}

\begin{document}
\makeatletter
\def\@copyrightspace{\relax}
\makeatother

\title{On the Reliability of Profile Matching\\ Across Large Online Social Networks}

\numberofauthors{4} 

\author{
\hspace{-10mm}
\alignauthor
Oana Goga\\
	\affaddr{MPI-SWS}
	\alignauthor
Patrick Loiseau\\
	\affaddr{EURECOM}
\alignauthor
Robin Sommer\\\
       \affaddr{ICSI}
       \and
\alignauthor Renata Teixeira\\
       \affaddr{Inria}
\alignauthor Krishna P. Gummadi\\
      \affaddr{MPI-SWS}
}

\maketitle

Matching  the profiles of a user across multiple online social networks brings opportunities for new services and applications as well as new insights on user online behavior, yet it raises serious privacy concerns. 
Prior literature has proposed methods to match profiles and showed that it is possible to do it accurately, but using evaluations that focused on sampled datasets only. 
In this paper, we study the extent to which we can \emph{reliably} match profiles \emph{in practice}, across real-world social networks, by exploiting \emph{public attributes}, i.e., information users publicly provide about themselves. 
Today's social networks have hundreds of millions of users, which brings completely new challenges as a reliable matching scheme must identify the correct matching profile out of the millions of possible profiles. 
We first define a set of properties for profile attributes--Availability, Consistency, non-Impersonability, and Discriminability (ACID)--that are both necessary and sufficient to determine the reliability of a matching scheme. 
Using these properties, we propose a method to evaluate the accuracy of matching schemes in real practical cases. Our results show that the accuracy in practice is significantly lower than the one reported in prior literature. 
When considering entire social networks, there is a non-negligible number of profiles that belong to different users but have similar attributes, which leads to many false matches.  
Our paper sheds light on the limits of matching profiles in the real world and illustrates the correct methodology to evaluate matching schemes  in realistic scenarios.

\section{Introduction}\label{sec:intro}

\noindent Internet users are increasingly revealing information about
different aspects of their personal life on different social
networking sites. Consequently, there is a growing interest in the
potential for aggregating user information across multiple sites, by
matching user accounts across the sites, to develop a more complete
profile of individual users than the profile provided by any single
site.  For instance, companies like PeekYou~\cite{peekyou} and
Spokeo~\cite{spokeo} offer {\em ``people search''} services that can
be used to retrieve publicly visible information about specific users
that is aggregated from across a multitude of websites. Some companies
are mining data posted by job applicants on different social
networking sites as part of background
checks~\cite{SocialIntelligence}, while others allow call centers to
pull up social profiles when their customers
call~\cite{salesforce}. The many applications of matching profiles
across social networking sites also raise many legitimate and serious
concerns about the privacy of users.  A debate on the relative merits
of leveraging profile matching techniques for specific applications is
out of the scope of this paper.

In this paper, our goal is to investigate the \emph{reliability} of
techniques for matching profiles across \emph{large real-world} online
social networks, such as Facebook and Twitter, using only {\it
  publicly} available profile attributes, such as names, usernames, location,
photos, and friends.
  Reliability refers to the extent to which different profiles belonging to the same user can be matched across
social networks, while avoiding mistakenly matching profiles belonging
to different users. Matching schemes need to be highly reliable
because incorrectly matched profiles communicate an inaccurate
portrait of a user and could have seriously negative consequences for
the user in many application scenarios. For example, Spokeo has been
recently sued over providing inaccurate information about a person
which caused ``actual harm'' to the person employment
prospects~\cite{spokeolawsuit}.
We focus on publicly available profile attributes because data
aggregators today can crawl and leverage such information for matching
profiles.

Recently, a number of schemes have been proposed for \matching
\profiles across different social networks
~\cite{Motoyama,PeritoCKM11,socialsearch,mywww,MishariT12,IofciuFAB11}
(we review them in~\xref{sec:existing_schemes}.) The potential of
these schemes to reliably match profiles in practice, however, has not
been {\it systematically} studied. Specifically, it is not clear how
or what properties of profile attributes affect the reliability of the
matching schemes. Furthermore, the training and testing datasets for
evaluating the matching schemes are often opportunistically generated
and they constitute only a small subset of all user profiles in social
networks.  It is unclear whether the reliability results obtained over
such datasets would hold over all user profiles in real-world social
networks, where there are orders of magnitude more non-matching
profiles than matching profiles (i.e., there is a huge class
imbalance).

Our first contribution lies in 
defining a set of properties for profile attributes--Availability,
Consistency, non-Impersonability, and Discriminability (ACID)--that are
both necessary and sufficient to determine the reliability of a
matching scheme (\xref{sec:acid}).
Analyzing the ACID properties of profile attributes reveals the
significant challenges associated with matching profiles reliably in
practice~(\xref{sec:acid_test}).
First, data in real-world social networks is often {\it noisy} --
users do not consistently provide the same information across
different sites. Second, with hundreds of millions of \profiles, there
is a non-trivial chance that there exist multiple \profiles with very
similar attributes (e.g., same name, same location) leading to
false matches. Finally, attackers create \profiles attempting to
impersonate other users, fundamentally limiting the reliability of any
profile matching scheme.

Another key contribution lies in our method for carefully selecting
the training and testing datasets for matching profiles~(\xref{sec:training}). When we
evaluate the main types of matching schemes in the literature (based
on binary classifiers) 
using a small random sample of Twitter and Facebook profiles (similar
to how these schemes were evaluated originally), the schemes achieve
over 90\% recall and 95\% precision
(\xref{sec:linker_small}). Unfortunately, when we evaluate these
schemes over datasets sampled carefully to preserve the reliability
that the schemes would have achieved over the larger datasets (full
Facebook database),
their performance drops significantly (\xref{sec:nary_linker}).  We
could obtain only a 19\% recall for a 95\% precision. 

We then investigate if we could improve the reliability of matching
schemes in scenarios where we know that there is \emph{at most one
  matching profile} (see~\xref{sec:3step}). In such scenarios, we propose a new matching scheme 
and show that it is indeed possible to improve the recall to 29\% at 95\% precision. This is still considerably lower
than the high recall (90\%) reported in the literature.

Thus, we discover a fundamental limitation in matching profiles across
existing social networks using public attributes. To further confirm the inherent limits of reliably matching profiles in practice, we compare the
reliability of automated matching schemes with that of human Amazon
Mechanical Turk (AMT) workers. Under similar conditions, AMT workers
are able to match only 40\% of the profiles with a 95\% precision.
Our analysis is the first to highlight that achieving high reliability
in \matching profiles across large real-world social networks comes at
a significant cost (in terms of reduced recall).

\section{Problem definition}
\label{sec:problem_definition}

\noindent In this section, we define the problem of matching profiles, we present the constraints we have to consider and discuss how we approach the problem.

 \vspace{2mm}
 \noindent \minisec{The profile matching problem}
 \noindent We consider that two profiles in two social networks match if they belong to/are managed by the same user. 
 The profile matching problem is: given a profile $a^{1}$ in one large social network $SN_1$, find all its \emph{\matchingprofiles} in another  large
social network $SN_2$, if at least one exists. We will denote by $a^{2}$ generic profiles in $SN_2$ and by $\hat{a}^2$ \matchingprofiles of $a^{1}$. For conciseness, we will also write $\match$ if $a^{2}$ is a \matchingprofile of $a^{1}$ and $\nonmatch$ otherwise. 


Note that we address here the problem of matching \emph{individual} profiles, which is  different from the problem of matching two entire social networks or databases. The difference is that we do not assume that we have access to all the data in  $SN_1$ but  only to one profile. 
For example, we cannot match profiles by exploiting patterns in the graph structure of $SN_1$ and $SN_2$, and we cannot optimize the matching of a profile in $SN_1$  based on the matchings of other profiles in $SN_1$. 
Thus, we cannot take advantage of some methods proposed for de-anonymizing social graphs~\cite{narayanan09osn,DBLP:journals/pvldb/KorulaL14} and entity matching~\cite{books/daglib/0030287}. 
 
Our problem formulation is motivated by practical scenarios.  There are many people search engines such as Spokeo that allow users to search for data about a particular person. These services gather data about a person by matching the profiles a person has on multiple social networks. 

We are particularly interested in two instantiations of the problem that are motivated by practical scenarios: (1) the \emph{generic  case} -- a profile can have multiple matching profiles in $SN_2$; and (2) the \emph{special case} -- a profile can have \emph{at most one} matching profile. This case is suited for matching social networks such as Facebook or LinkedIn that enforce users to have only one profile. 

 \vspace{2mm}
 \noindent \minisec{Features}
\noindent In this paper, we investigate the extent to which we can match profiles by exploiting the \emph{attributes} users publicly provide in their profiles such as their \emph{real names, screen names}  (aka. username -- name that appears in the URL of the profile), \emph{location, profile photos,} and \emph{friends}.  
Using this information we can ideally match any person that maintains the same persona on different social networks. Also, we choose these attributes because they are essential to find people online and they are present and usually remain public across different social networks even if users make all their other content, such as their posts and photos, private. 
For profile $a^{1}$ (resp. $a^{2}$), we denote by $v^{1}$ (resp. $v^{2}$) the value of a considered attribute. From attribute values, we define a \emph{feature} as the similarity between the values of profiles in $SN_1$ and $SN_2$: $s(v^{1},v^{2})$.

 \vspace{2mm}
  \noindent \minisec{Matching scheme as a binary classifier}
\noindent Most previous works solved the matching problem by building binary classifiers that, given two profiles $a^{1}$ and $a^{2}$, determine whether $a^{1}$ and $a^{2}$ are \matching or not~\cite{MishariT12,5636108,5272173,conf/socialcom/PeledFRE13,socialsearch,Motoyama,jain:Iseekfbme:2013:wole,malhotra1}.
The binary classifier takes as input a feature vector $f(a^{1}, a^{2})$ that captures the similarity
between each attribute of a pair of profiles $(a^{1},a^{2})$; and then outputs the probability $p$ of $a^{1}$ and
$a^{2}$ to match. By selecting a cut-off threshold for $p$  the classifier  returns 1 (i.e.,
\matchingprofiles) if $p$ is larger than the threshold; and 0
otherwise. We say that a matching scheme outputs a \emph{true match} when the matched profiles  belong to the same user and outputs a \emph{false match} when the matched profiles  belong to different users. The threshold's choice constitutes the standard tradeoff between increasing the number of true match and decreasing the number of false matches.

This solution works well for the generic case of our matching problem. Given a profile $a^{1}$, we can use the binary classifier to check, for every pair of profiles $(a^{1},a^{2})$ such that $a^{2} \in SN_2$,  whether it is matching or not. We can then output any profile 
$a^{2}$ that the binary classifier declares as \matching.  In this paper, we test such approach when we represent $(a^{1},a^{2})$ with five features, each corresponding to the similarity score between $a^{1}$ and $a^{2}$ for each of the five profile attributes: real name, screen name, location, photo, and friends. 



For the special case of our matching problem, the previous approach is vulnerable to output many false matches.  For this case, instead of independently judging whether \emph{each} pair $(a^{1},a^{2})$ is a match or not, we can compare (for a given $a^{1}$) the probabilities $p$ for \emph{all} pairs $(a^{1},a^{2})$ to judge which profile is most likely the \matchingprofile of $a^{1}$. We discuss this case in more detail in~\xref{sec:3step}. 


 \vspace{2mm}
 \noindent \minisec{Reliability of a  profile matching scheme} 
\label{sec:evaluation}
\noindent  In this paper our focus is on the reliability of matching schemes.  A \emph{reliable matching scheme}
should ensure that the \profile it finds indeed matches with 
high probability, i.e., the matching scheme does not have many false matches.  
If there is no clear matching profile in $SN_2$ for $a^1$, then the scheme should return nothing.

Many previous studies used the true and the false positive rate to evaluate their matching schemes. The true positive rate is the percentage of matching profiles that are identified, while the false positive rate is the percentage of non-matching profiles that are false matches. The goal is to have a high true positive rate and a low false positive rate. 
These metrics are, however, a misleading indicator of the reliability of a matching scheme because they are not suited for scenarios with high class imbalance, i.e., the number of matching profiles is much lower than the number of non-matching profiles. 
%
For example,  a matching scheme with a 90\% true positive rate for a 1\% false positive rate might seem reliable, however, if we use it in a scenario where we have 1,000 matching and 999,000 non-matching profiles,  the matching scheme would output 900 true matches and 9,990 false matches, which is clearly unreliable. In real-world social networks, the class imbalance is even higher (e.g., for each matching profile we have over 1 billion non-matching profiles in Facebook) thus the scheme would output even more false matches. 

This paper argues that better metrics to evaluate the reliability of a matching scheme are the precision and recall. The recall is the percentage of matching profiles that are identified, while the precision the percentage of all pairs returned by the matching scheme which are true matches. The goal is to have a high recall and a high precision.  In the previous example, we would have
90\% recall for a 8\% precision, which shows the low reliability of the scheme (out of all matched profiles only 8\% are true matches).
Thus, the best way to show the reliability of a matching scheme is to evaluate its precision and recall with realistic class imbalance. In the rest of the paper, by \emph{reliable} we mean a precision higher than 95\%.

\section{The ACID framework}
  \label{sec:acid}

\noindent   The natural question that arises when investigating the reliability of matching schemes is: what does the reliability depends on?  Undoubtedly,  the reliability depends on the attributes we consider for matching and on their properties. Thus, given an attribute, what properties should the attribute have in order to enable a reliable profile matching?
We propose a set of four properties to help capture the quality of different attributes to match profiles: {\it Availability, Consistency, non-Impersonability, and Discriminability (ACID)}.

\noindent \textbf{Availability:}  At first, to enable finding the \matchingprofile, an attribute should have its value available in both social networks. 
If only 5\% of users provided information about their ``age'' across two sites, then ``age'' has limited utility in matching profiles. 
To formalize this notion, we model the attribute values of $a^{1}$ and each $a^{2} \in SN_2$ as random variables and we define the availability of an attribute as:  
$$A = Pr \left(v^{1} \textrm{ and }   v^{2} \textrm{ available} \big| \match\right).$$ 


\noindent \textbf{Consistency:} It is crucial that the selected attribute is
  consistent across \matchingprofiles, i.e., users provide
  the same or similar attribute values across the different profiles they manage. 
 Formally, we define the consistency of an attribute as: 
 $$C = Pr\left(s(v^{1}, v^{2}) > th \big| \match, v^{1} \textrm{ and } v^{2} \textrm{ available}\right),$$ where $th$ is a threshold parameter.  

\noindent \textbf{non-Impersonability:} If an attribute can be easily impersonated,
  i.e., faked, then attackers can compromise the reliability of the
  matching by creating fake profiles that appear to be matching with
  the victim's profiles on other sites. Some public attributes like
  ``name'' and ``profile photo'' are easier to copy 
  than others such as ``friends''.
  To formalize this notion, we introduce the notation $\impers$ to denote that profile $a^{2}$ has been created by an attacker impersonating profile $a^{1}$. We denote the probability that there exists at least one profile $a^{2}$ impersonating $a^{1}$ by $p_{I} = Pr (a^{1} \textrm{ is impersonated})$ and the probability that there is no profile impersonating $a^{1}$ by $p_{nI} = 1 - p_{I}$. The difficulty to manipulate an attribute is characterized by its non-Impersonability defined as: 
  $$
  nI = Pr \Big(\max_{a^{2}: \impers} s(v^{1},v^{2}) <  th \Big).
  $$



\noindent \textbf{Discriminability:} 
Even without impersonations, in order to enable finding the \matchingprofile, an attribute needs to uniquely identify a profile in $SN_2$. 
 A highly discriminating attribute would have a unique and different value for each profile, while a less discriminating attribute would have similar values for many profiles. For example, ``name'' is likely to be more discriminating than  ``gender''. 
  Formally, we define the discriminability of an attribute as:  
$$D = Pr\Big(\max_{a^{2}: \nonmatch} s(v^{1},v^{2}) <  th  \big| a^{1} \textrm{ not impersonated}\Big).$$  
  In practice, it is impossible to estimate $D$ unless we are able to identify impersonating profiles. Instead, we estimate: 
$$\tilde{D} = Pr\Big(\max_{a^{2}: \nonmatch} s(v^{1},v^{2}) <  th\Big).$$
$\tilde{D}$ represents the ``effective discriminability'' taking into account possible impersonations.     
Since impersonators create non-matching profiles as similar as possible to the original profile, it is reasonable to assume that $\tilde{D}\le D$. 
Moreover, by application of Bayes formula, we can show that $D \le \tilde{D}/p_{nI}$ so that, if $p_{I}$ is not too large, $\tilde{D}$ gives a good estimate of $D$. If we assume that the impersonating profiles are independent from the other non-matching profiles, we can also prove that $\tilde{D} = D \cdot (p_{nI} + nI \cdot p_{I})$. This clearly shows that $\tilde{D}$ is close to $D$ if either the attribute is hard to impersonate ($nI$ close to one) or the proportion of impersonator is small ($p_I$ small).

  
  



The ACID properties are clear and intuitive properties that help understand the potential of an attribute to perform reliable matching, as the following theorem formalizes.~\footnote{\small{The proof can be found in the Appendix~\ref{ap:proof}.}}

\newtheorem{mydef}{Theorem}
 \begin{mydef}
 \vspace{-2mm}
\label{theorem}
Consider a classifier based on a given attribute that classifies as matching profiles if $s(v^{1}, v^{2})>th$. The performance of the classifier is characterized by the following results. 
\vspace{-2mm}
\begin{trivlist}
\item[\hspace{1mm} \textit{(i)}] We have\\[-6mm] $$recall = C \cdot A.$$ 
\item[\hspace{1mm}  \textit{(ii)}] Assume that, for each profile $a^{1}\in SN_1$, there is at most one matching profile in $SN_2$. Then, $$precision \le \frac{recall}{recall + 1 - \tilde{D}}.$$
\item[\hspace{1mm}  \textit{(iii)}] Assume that $p_{I}>0$. Then, $precision = recall =1$ iif $A = C = nI = D = 1.$
\end{trivlist}
\vspace{-4mm}
\end{mydef}
%

In Theorem~\ref{theorem}, the threshold parameter $th$ must be the same as the one in the definitions of $C$, $nI$ and $D$. 
Theorem~\ref{theorem}-\textit{(i)} shows that the classifier's recall is simply the product of consistency and availability. Theorem~\ref{theorem}-\textit{(ii)} gives a simple upper bound of the precision as a function of the effective discriminability (which itself is a function of the discriminability and of the impersonability, see above). This upper bound gives a good order of magnitude for the precision; moreover, for high precision (which is what we aim), given the small number of false positives, the true precision should be close to the bound. Finally, Theorem~\ref{theorem}-\textit{(iii)} confirms that a high value of all four ACID properties is \emph{necessary} and \emph{sufficient} to obtain high precision \emph{and} recall.

Properties $A$, $C$ and $nI$ are independent of the network scale, however,  the discriminability very largely depends on the network scale since having more non-matching pairs decreases the probability that none of them has a high similarity score. This implies that we must estimate the precision and the recall of a matching scheme using datasets that accurately capture the ACID properties of profile attributes of the entire social network. Otherwise, the precision and the recall will be incorrect.

In practice different attributes satisfy the properties to different extent and the challenge is to combine different attributes with imperfect properties to achieve a reliable matching. The next section analyzes the ground truth for several large social networks to understand the limits of matching profiles across different sites. 

\section{Limits of matching profiles}
\label{sec:acid_test}


\noindent To understand the limits of matching profiles, we analyze the ACID properties of profile attributes (screen name, real name, location, profile photo, and friends) across six popular social networks (Facebook, Twitter, Google+, LinkedIn, Flickr, and MySpace). 
First we present our method to gather ground truth of matching profiles and we then analyze each property separately.

\subsection{Ground truth of matching profiles}
\label{sec:gt}
\noindent 
Gathering ground truth of
\matchingprofiles spanning multiple social networks is challenging and many previous works manually labeled profiles~\cite{Liu:2013:WNU:2433396.2433457,5272173,conf/socialcom/PeledFRE13}. Below we describe two automatic methods that we used to obtain our ground truth. 

We  first obtained ground truth data by exploiting ``Friend Finder''
mechanisms on many social networks that allow a user to find
her friends by their emails. We used a list of
email addresses collected by colleagues for an earlier study analyzing
spam email~\cite{Kreibich:2009:SIL:1855676.1855680}.\footnote{\small{The local IRB approved the collection.}}  These
email addresses were collected on a machine instructed to send spam by
a large bot network.  Since
spammers target the public at large we believe that this list of
emails catch a representative set of users. To combat abuse, some social networks limit the
number of queries one can make with their ``Friend Finder'' mechanism
and employ techniques to make an automated matching of an email to a
profile ID impossible. Hence, we were only able to collect the
email-to-profile ID matching for Twitter, Facebook, LinkedIn and Flickr.
Table~\ref{tab:data:GT} summarizes the number of \matchingprofiles we
obtained using the Friend Finder mechanism (\datasetFF).\footnote{\small{To test the representativeness  of \datasetFF, we compare the distribution of properties  such as account creation date, number of followers, and number of tweets of Twitter 
profiles in our dataset with the same properties of random Twitter profiles. We found that the pairs of distributions for each property matched fairly closely.}}

Some previous works obtained ground truth from users that willingly provide links to their profiles in different social networks.  Such users might not represent users in general because they want their profiles to be linked and probably expend the effort to keep their profiles synced.  
%
%
To be able to compare our results against previous works we collected \datasetG (see Table~\ref{tab:data:GT}) by exploiting the fact that Google+ allows users to explicitly
list their profiles in \emph{other} social networks on their profile pages.
Due to space constraints, for the rest of the paper, we show by default the results for profiles in \datasetFF and occasionally, for comparison, we show   the results are for \datasetG.




\begin{table}[t]
\scriptsize{
\caption{\small{Number of ground truth \matchingprofiles obtained with Friend Finder (\datasetFF) and Google+ (\datasetG) for different combinations of social networks.}}
\vspace{-3mm}
\label{table:avbw}
{\centering
\begin{tabular}{l|c|c}
 & \textbf{\datasetFF} & \textbf{\datasetG} \\
\hline
\hline
\textbf{\textsc{Twitter - Facebook}} & 4,182 & 76,332  \\
\textbf{\textsc{LinkedIn - Facebook}} & 2,561 & 20,145 \\
\textbf{\textsc{Twitter - Flickr}} & 18,953 & 35,208 \\
\textbf{\textsc{LinkedIn - Twitter}} & 2,515 & 20,439 \\
\end{tabular}

}
\label{tab:data:GT}
\vspace{-1\baselineskip}
}
\end{table}

\subsection{Attribute availability}
\label{sec:avail}
\noindent 
The availability of attributes depends on the social network, for example Twitter does not ask users about their age while Facebook does. The availability also depends on whether users  choose to input the information and make it public.  Users might choose to let their location public on Twitter while make it private on Facebook.  

Table~\ref{tab:avail} shows the breakdown of attribute availability per social network and pairs of social networks. The availability per social network characterizes the behavior of users, while the availability for pairs of social networks corresponds to the definition of $A$ in~\xref{sec:acid}.

\begin{table}[t]
\scriptsize{
{\centering
\caption{\small{Availability of attributes for \datasetFF. \\ \uline{Legend:} Tw~=~Twitter, Fb~=~Facebook, Fl~=~Flickr, Lnk~=~LinkedIn.}}
\vspace{-3mm}
\setlength{\tabcolsep}{2pt}
\begin{tabular}{l|c|c|c|c|c}
&  \textbf{Screen Name}   & \textbf{Real Name} & \textbf{Profile Photo} & \textbf{Location} & \textbf{Friends} \\
\hline
\hline
\textbf{\textsc{Tw}} & 100\% & 100\% & 69\% &  54\% & 86\% \\
\textbf{\textsc{Fb}}  & 100\% & 100\% & 98\% &52\%  & 60\% \\
\textbf{\textsc{Fl}}  & 100\% & 30\% &29\%& 11\% & 40\% \\
\textbf{\textsc{Lnk}}  & 100\% & 100\% & 57\% &99\% & 0\%\\
\hline
\hline
\textbf{\textsc{Fb - Tw}}  & 100\% & 100\% & 69\% & 30\% & 43\%\\
\textbf{\textsc{Fb - Lnk}} & 100\% & 100\% & 56\% & 54\% & 0\%\\
\textbf{\textsc{Tw - Fl}} & 100\% & 30\% & 24\% & 8\% & 32\% \\
\textbf{\textsc{Lnk- Tw}}  & 100\% & 100\% & 44\% & 54\% & 0\%\\
\end{tabular}
\label{tab:avail}

}
\vspace{-2\baselineskip}
}
\end{table}%

First, we find that the availability of the attributes varies
considerably across the different social networks. For
example, users are more likely to provide their location information
on LinkedIn than they are on Facebook or Twitter. The differences in
availability are presumably due to the different ways in which users
use these sites. For our purposes, it highlights the additional
information one could learn about a user by linking her profiles on
different sites. 

Second, we find that screen name and real name are
considerably more available than location or
friends.  
However, the availability of the less available attributes is not negligible -- for example, location and
friends are available for more than 30\% of matching profiles in Twitter and Facebook. 

Third, when we compare the availability using \datasetFF and \datasetG (not shown), we observe that the availability of attributes for profiles in the \datasetFF is much lower than the availability for profiles in the \datasetG (e.g., profile photo is available for only 69\% of Twitter users in \datasetFF while it is available for 96\% of users \datasetG).\footnote{\small{For more results on \datasetG, we refer the reader to~\cite{Goga}.}} 
Thus, users in \datasetG are more likely to complete their profiles and consequently there is a higher bound on the recall to match them.


\subsection{Attribute consistency}\label{sec:features}
\label{sec:consistency}

\noindent We now study the extent to which users
provide consistent attribute values for their profiles on different
social networks. Some users deliberately provide different
attribute values either out of concerns for privacy or out of a desire
to assume online personas different from their offline persona. It
would be very hard to match \profiles of such users by
exploiting their public attributes. 

Other users may input slightly different values for an attribute across sites. For example, a user might
specify her work place as International Business Machines on one site and
International Business Machines Corporation on another site. 

 \vspace{2mm}
 \noindent \minisec{Similarity metrics for profile attributes}
\noindent We borrow a set of standard metrics from prior work in
security, information retrieval, and vision communities to compute
similarity between the values of  attributes:  the Jaro distance~\cite{Cohen03acomparison} to measure the similarity between names and screen names; the geodesic distance to measure the similarity between locations; the phash~\cite{phash} and SIFT~\cite{SIFT} algorithms to detect whether two photos are the same; and the number of common friends between two profiles. Please check our Appendix~\ref{ap:sim} for a full description of these metrics.

 \vspace{2mm}
 \noindent \minisec{Similarity thresholds for attribute consistency}
\noindent 
Clearly the more similar two values of an attribute, the
greater the chance that the values are consistent, i.e., they refer to
the same entity, be it a name or photo or location. 
Here, we want to show consistency results for a ``reasonable'' threshold beyond which we can declare with high confidence that the attribute values are consistent (irrespective of the tradeoff between consistency and discriminability in~\xref{sec:acid}). The best to judge whether two attribute values are consistent are humans.
%
%
Thus, we gathered ground truth 
data by asking Amazon Mechanical Turk (AMT) users to evaluate whether pairs of attribute
values are consistent or not.  We randomly select 100 pairs each of
matching and \nonmatching Twitter and Facebook profiles from
\datasetFF and asked AMT users to annotate which attribute values are consistent and which are not. We followed the guidelines to ensure 
good quality results from AMT workers~\cite{AMT_guidelines}.  

For each attribute, we leverage the AMT experiment to select the similarity thresholds to declare two values as consistent. 
Specifically, we select similarity thresholds, such that more than 90\% of the consistent values, as identified by AMT workers, and less than 10\% of the inconsistent values have high similarities.
Note that, we only use these thresholds to evaluate whether attribute values in \matchingprofiles are consistent and we do not use them to actually match profiles. Thus, while it is important that the majority of  consistent values pass the threshold,  it is not critical if some inconsistent  values also pass the threshold.
Incidentally, this experiment also shows that the similarity metrics we choose are consistent with what humans think it is similar.
Note that, it is unpractical to use AMT workers to estimate the threshold for friends, thus we manually choose it  be at least two friends in common to avoid noise.

\begin{table}[t]
\scriptsize{

{\centering
\setlength{\tabcolsep}{2pt}
\caption{\small{Consistency of attributes for users in \datasetFF ; $\dagger$ in parenthesis, the consistency only when information is available in both social networks.}}

\vspace{-3mm}
\begin{tabular}{l||c|c|c|c|c}
  & \textbf{Screen Name}  & \textbf{Real Name} & \textbf{Location} & \textbf{Profile Photo}  &  \textbf{Friends}  \\
\hline
\hline 
Fb - Tw  & 38\% & 77\% & 23\% (77\%$\dagger$)  & 8\% (12\%) & 34\% (79\%) \\ 
Fb - Lnk  & 71\%  & 97\%  & 44\%  (83\%) & 11\% (23\%) & 0\%  \\ 
Tw - Fl & 40\%  & 25\%  (84\%)  & 5\% (67\%) & 5\% (22\%)  & 13\% (42\%)\\ 
Tw - Lnk & 36\%  & 83\% & 39\% (71\%) & 13\% (31\%) & 0\%\\ 

\end{tabular}
\label{tab:consistency}

}
\vspace{-2\baselineskip}
}
\end{table}%

 \vspace{2mm}
 \noindent \minisec{Attribute consistency in \matchingprofiles}\label{sec:consistency}
\label{sec:prop}
\noindent Table~\ref{tab:consistency} shows the proportion  of users who provide consistent values for an attribute in a pair of social networks out of all users.This proportion corresponds to the recall we can achieve using the attribute given the threshold used, as shown in the previous section.  In parenthesis, we also provide the equivalent proportion of users with consistent values only when the attribute value is available in both social networks (corresponding to the definition of $C$). This proportion better illustrates  how likely users are to provide consistent values, i.e., shows the users's attempt to maintain synched profiles. 


First, we find that a large fraction of users provides similar {\it real names} across different social networks. Put differently, most users are not attempting to maintain distinct personas on different sites. This trend bodes well
for our ability to match the profiles of a user. 

Second, we  computed the percentage of \matchingprofiles in Twitter and Facebook
for which all public attributes in Table~\ref{tab:consistency} are
inconsistent. We find that there are 7\% of such users. These
users are likely assuming different personas on different sites and it
is very hard, if not impossible, to match their profiles using only the
public attributes that we consider in this paper. Thus, we can at most
hope to match profiles for 93\% of users. This percentage 
represents an upper bound on the recall for matching profiles based on public attributes.


Third, the consistency differs between different social networks. Twitter and Facebook have one of the lowest consistency for each attribute while Facebook and LinkedIn have the highest consistency. Thus, users are more likely to maintain synched profiles across Facebook and LinkedIn than other pairs of social networks.

\subsection{Attribute discriminability}
\label{sec:discr}

%
%

\begin{figure*}[!htbp]
\vspace{-2mm}
{\centering
\subfloat[\small{Real name}]{\label{fig:cdf_new_m_nm_n}\includegraphics[width=0.20\textwidth,height=2.5cm]{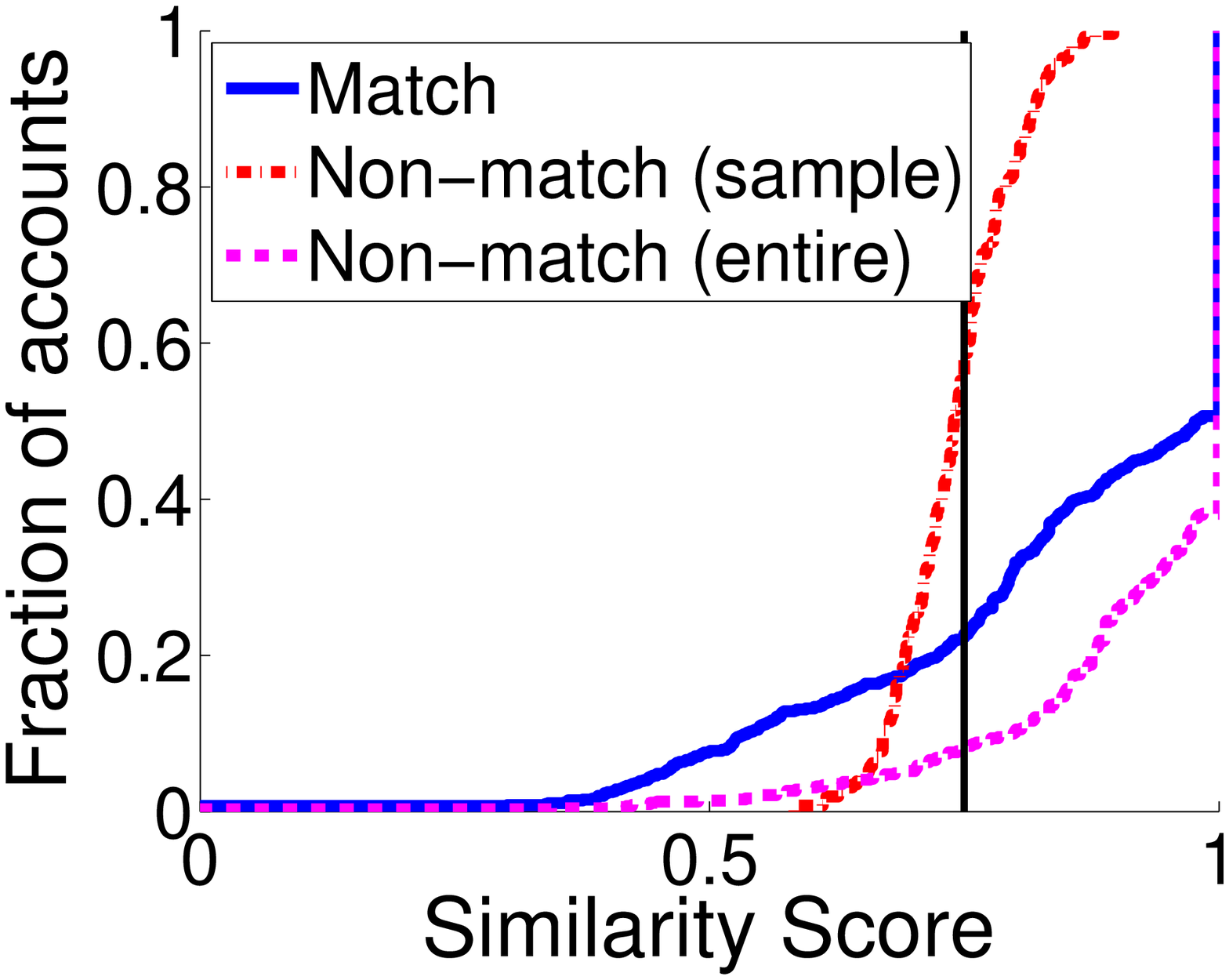}}
\subfloat[\small{Screen name}]{\label{fig:cdf_new_m_nm_u}\includegraphics[width=0.20\textwidth,height=2.5cm]{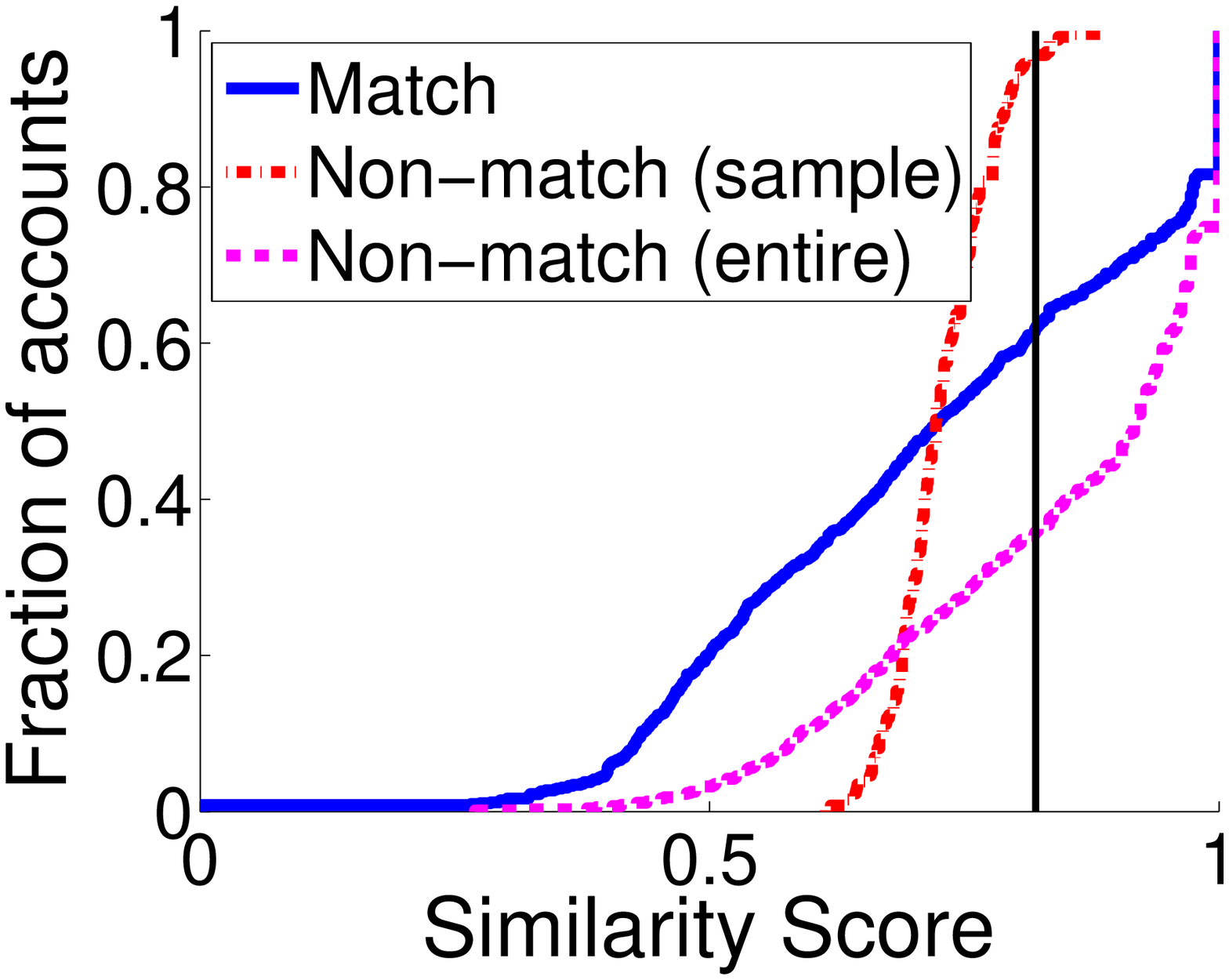}}
\subfloat[\small{Photo (SIFT)}]{\label{fig:cdf_new_m_nm_s}\includegraphics[width=0.20\textwidth,height=2.5cm]{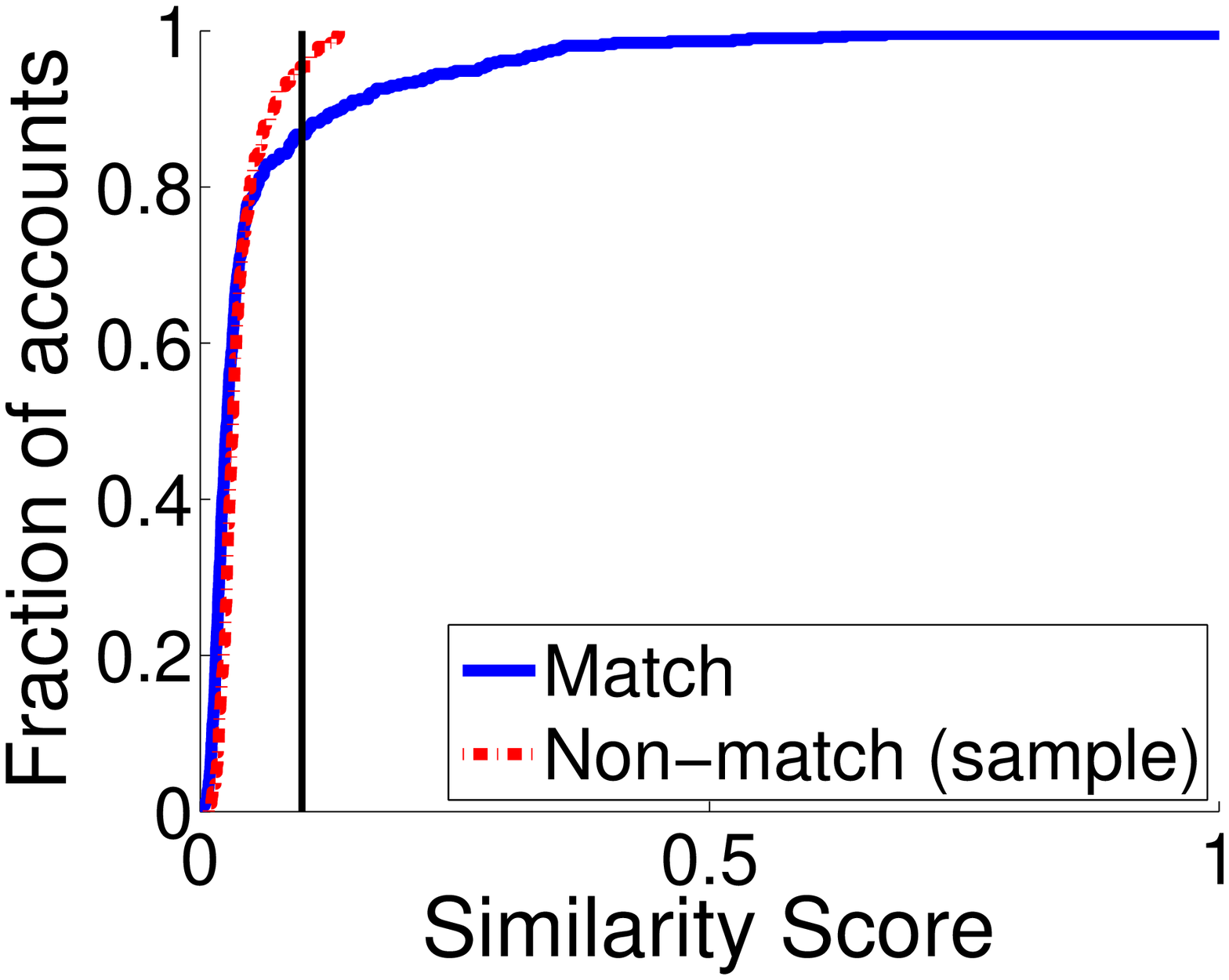}}
\subfloat[\small{Location}]{\label{fig:cdf_new_m_nm_l}\includegraphics[width=0.20\textwidth,height=2.5cm]{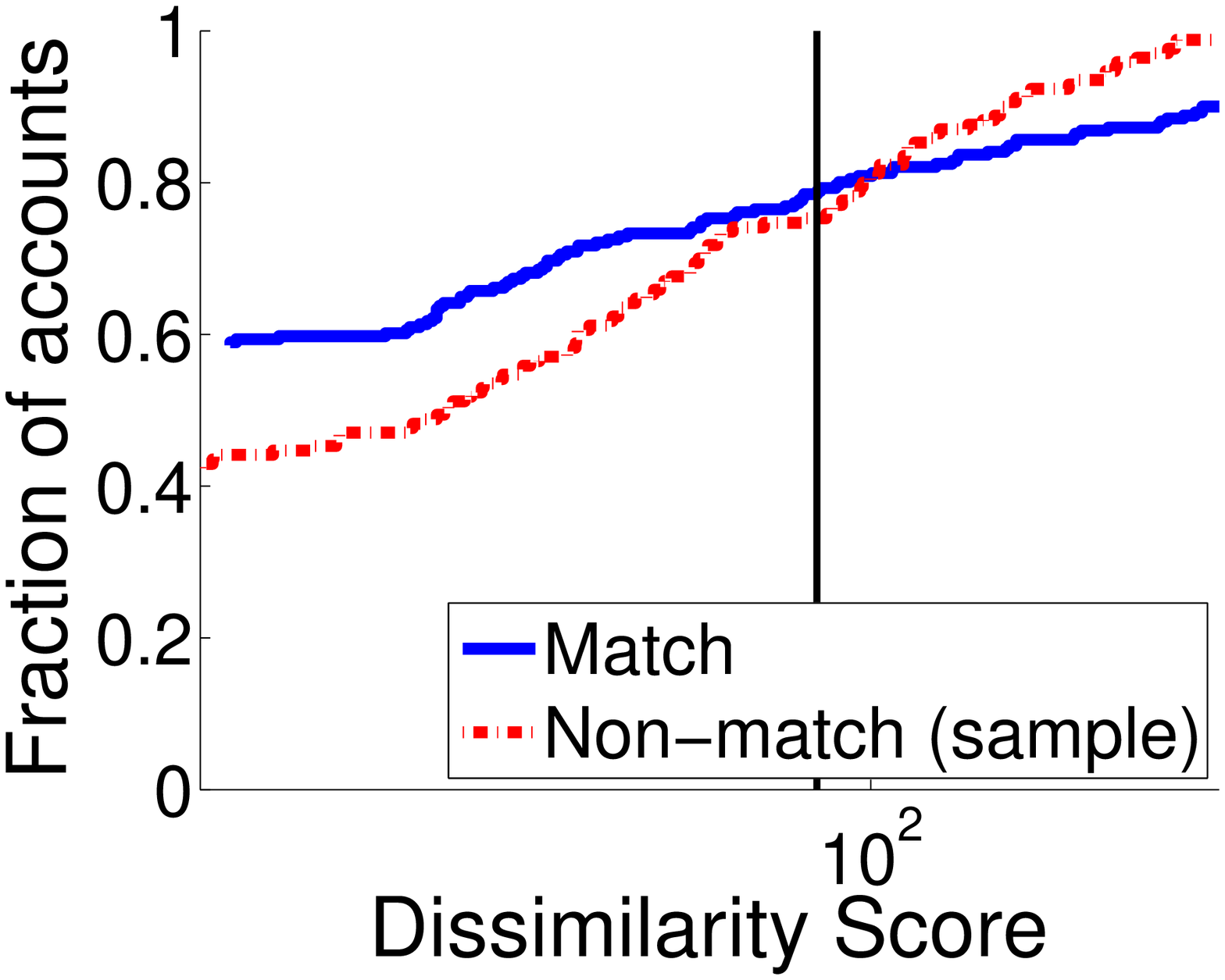}}
\subfloat[\small{Friends}]{\label{fig:cdf_new_m_nm_f}\includegraphics[width=0.20\textwidth,height=2.5cm]{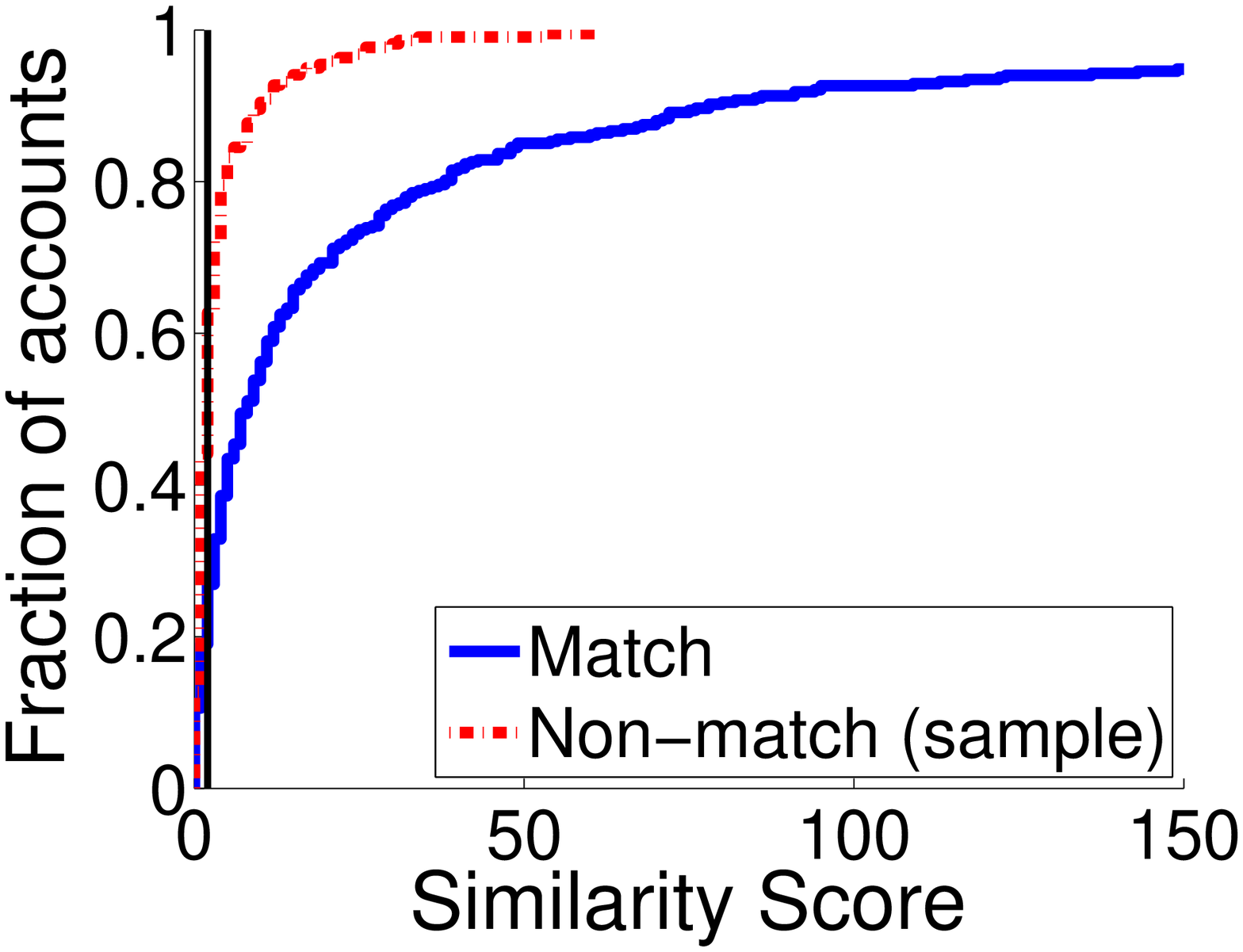}}


\vspace{-2mm}
\caption{\small{CDF of similarity scores for the matching Facebook profiles and for the most similar non-matching Facebook profile  in \datasetFF(sample) and the entire  Facebook (entire).}}
\label{fig:cdf_new_m_nm}
}
\vspace{-1.5\baselineskip}
\end{figure*}

\noindent The previous section showed that a large fraction of  users have consistent attribute values between their profiles. However, the number of profiles that we can match reliably is smaller because attribute values might not uniquely identify a single person.

To evaluate the discriminability of attributes, for each Twitter profile we compare the similarity of the matching Facebook profile with  the similarity of the most similar non-matching Facebook profile. 
Figure~\ref{fig:cdf_new_m_nm} shows the CDF of similarity scores in \datasetFF (sample). Zero means no similarity and
one means perfect similarity; except for location, where zero means perfect similarity
because it corresponds to the distance between locations. 
The vertical lines represent the similarity thresholds for consistent attribute values used in the previous subsection. Given a threshold, we have perfect discriminability if there are no non-matching profiles with higher similarities. 
Concretely, for a given similarity threshold (x value), the y value for the distribution for the most similar non-matching profile represents an estimate of the (effective) discriminability $\tilde{D}$, whereas the y value for the distribution of matching profiles represents the complementary of the recall $1-C\cdot A$.


For the \emph{real name} and \emph{screen name} we see a clear
distinction between distributions of  matching and \nonmatching profiles in Figure~\ref{fig:cdf_new_m_nm}. The highest similarity of non-matching profiles is around 0.75 while a number matching profiles have similarities around 1.  This suggests that these attributes have a high discriminability. 
For \emph{photo}, the two distributions are generally similar. The photo does not appear to have a very good overall discriminability because there are not many Facebook \matchingprofiles that use the same profile photo with the Twitter profile. However, for similarities large than 0.10,  when the profile photos are consistent, there are not many non-matching profiles. 
As expected, the \emph{location} does not have a good discriminability; even in a small dataset there are Facebook non-matching profiles with the same location as the Twitter profile. 
Finally,  \emph{friends} have a good discriminability between matching and \nonmatching profiles, i.e., it is uncommon to have non-matching profiles with many common friends.

We do not have access to the whole Facebook dataset to evaluate the discriminability of all attributes over an entire social network, however, we exploit the Facebook Graph Search to estimate the discriminability of real names and screen names. For each Twitter profile we use Facebook Graph Search to retrieve all the profiles with the same or similar names and screen names. This procedure samples the non-matching profiles with the highest similarity; therefore it preserves the discriminability of the entire social network.
%
Figure~\ref{fig:cdf_new_m_nm_n} and~\ref{fig:cdf_new_m_nm_u} also presents the discriminability of real names and screen names over the entire Facebook (entire). As expected, the CDF of similarity score for non-matching profiles is much lower than it was at small scale.
Furthermore, for 60\% of the Twitter profiles, there is a non-matching Facebook profile that has \emph{exactly} the same real name and for 25\% exactly the same screen name. Even worse, the CDFs of Figure~\ref{fig:cdf_new_m_nm_n} for non-matching profiles are even below the CDFs for matching profiles which means that in many cases there are non-matching profiles that have even more similar names  with the Twitter profile than the matching profile.  These results show that names and screen names are actually not so discriminating in practice  and consequently shows the difficulty of reliably finding the matching profile in real-world social network. This also shows the risk of evaluating matching scheme over a sampled dataset because attributes have a much higher discriminability than over entire social networks.

\subsection{Attribute  impersonability}
\noindent In most social networks a user is not required to prove that her online identity matches her offline person. 
Since there is a lot of  personal data  publicly available,  it is very easy for attackers to create fake profiles that impersonate honest  users.  
Because such attacks could be a very big source of unreliability for matching schemes, we show evidence that such attacks indeed exist and they are more frequent than previously assumed. 

 To search for potential cases of impersonation we start with an initial set of 1.4 million random profiles in Twitter.  
 We find that, strikingly, a large fraction of profiles could be \emph{potential} victims of impersonation attacks:  18,662 Twitter profiles have at least another Twitter profile with consistent profile attributes. This gives a rough estimate of $p_{I}$ of 1\%.
 It is beyond the scope of this paper to thoroughly investigate such attacks but in~\xref{sec:3step} we propose a way to make matching schemes less vulnerable to impersonation attacks.

%
%


\section{Training \& Testing Matching \\  Schemes}\label{sec:data}
\label{sec:training}
\noindent In this section, we focus our attention on the datasets used
to train and evaluate (test) matching schemes. 
To estimate well the precision and recall in practice, we should test for each profile $a^{1}$ the accuracy of finding the \matchingprofiles $\hat{a}^{2}$ out of all the
profiles $a^{2} \in SN_2$.  If we consider large social networks like
Facebook, Twitter, or Google+, $SN_{2}$ has hundreds of millions of
profiles.   Obtaining such complete datasets is impractical, thus, we have to sample a number of profiles in the network.

Most previous studies sampled datasets by picking matching and non-matching profiles at random.  
Such random sampling fails to capture the precision and recall of matching schemes in practice because  it severely over-estimates the discriminability of attributes found in the original social network (as seen in~\xref{sec:discr}) and therefore it severely over-estimates precision. 
To estimate well the reliability of a matching scheme in practice, the sampled dataset needs to preserve the precision and recall of the original social network at least for high values of precision.
The key to ensure this is to sample \emph{all potential} false matches, i.e., all profiles that could be mistakenly matched by the matching scheme. 
Thus, we build two datasets: (1) a \emph{reliability non-preserving sampled dataset} for
comparison with previous techniques (as this is the standard
evaluation method); and (2) a \emph{reliability preserving sampled dataset} that strives  to capture all possible false matches in a social network to better estimate the reliability of matching schemes in practice. 

We generate a reliability preserving sampled dataset for
matching Twitter and Facebook. Although building such
dataset for other social networks is possible, the process is
strenuous. Instead, we take two of the most popular
social networks to show the limitations for matching
profiles across real-world social networks.

 \vspace{2mm}
 \noindent \minisec{Reliability-non-preserving sampling}
\noindent 
We randomly sample 850 matching Twitter-Facebook profiles from \datasetFF 
and we use them to build 722,500  pairwise
combinations of Twitter-Facebook profiles (850 positive  and 721,650 negative examples). We call the resulting dataset the \smallscale.
The \smallscale dataset preserves the availability and consistency of attributes in the original social network, but it does not preserve the discriminability and non-impersonability. Thus, the dataset does not preserve the precision of the original social network.  Note that,  datasets such as \datasetG, which have been used in previous work, do not even preserve the availability and consistency of attributes because they are biased towards a particular kind of users (as seen in~\xref{sec:avail}); hence they do not preserve recall.


%
%


 \vspace{2mm}
 \noindent \minisec{Reliability-preserving sampling}
\noindent To preserve the reliability over the original social network, our sampling strategy is to sample non-matching profiles that have a reasonably high similarity to $a^{1}$ and ignore non-matching profiles that have a very small chance of matching. We note the set of most similar profiles to $a^{1}$ in $SN_2$  as $C(a^{1}) \subset SN_2$. A  comprehensive $C(a^{1})$ includes all the Facebook profiles which could be potential false matches. 

Given that our analysis
in~\xref{sec:consistency} shows that  most Twitter-Facebook
\matchingprofiles have consistent real names or 
screen names, we hope to build a comprehensive $C(a^{1})$ by
exploiting the Facebook search API, which allows searching for people by name. 
%
For each Twitter profile, $a^{1}$ (we sample the same Twitter profiles from \smallscale), we generate $C(a^{1})$ using the Facebook search API to find
profiles with the same or similar real name or screen name as $a^{1}$.
The resulting dataset, which we call \largescale, contains over 270,000
combinations of profiles $(a^{1},a^{2})$ where  $a^{2} \in C(a^{1})$. Thus, for each Twitter profile the dataset contains in average 320 Facebook profiles with similar names.

Our analysis shows that the \matchingprofile of $a^{1}$ is in $C(a^{1})$ (i.e., $\hat{a}^{2} \in C(a^{1})$) for 70\% of Twitter profiles.
This implies that for 70\% of cases we selected at least all non-matching profiles with higher name similarity than the matching profile.  Additionally, the median similarity of the least similar real name in $C(a^{1})$ is 0.5, while the median similarity of matching profiles is 0.97.  This means that we also catch many Facebook profiles with lower name similarity than the matching profiles.  Thus, the only possible false matches that we miss are the ones that have very different names.  

Note that the reliability preserving sampling does not sample the matching profile when there is little chance for it to match (in 30\% of the cases). We actually tried to train and test matching schemes with or without including the unsampled matching profiles in the \largescale and the reliability did not differ significantly.

Our sampling strategy ensures that the discriminability and impersonability of real names and screen names found in the real-world datasets are preserved.  It might over-estimate, however, the discriminability of location, friends, and profile photos since we do not sample in $C(a^{1})$ profiles with similar location, friends or photos if they do not also have similar names or screen names. 
Evaluating matching schemes over $C(a^{1})$ rather than all $SN_2$ could lead to an under-estimation of 
the false matches. Thus, the precision we obtain over this dataset is an upper bound on the precision in practice.  This implies that the limits of reliably matching schemes in practice can only be worse than what we show in this paper. 
We believe, however, that our sampling strategy gives a very good idea of the precision and recall in the real-world datasets because there will be very few false matches (if any) with very dissimilar names even if they have similar location, photo or friends. 

Another limitation of the dataset is that it does not contain cases where a profile $a^{1}$ has multiple  matching profiles in $SN_2$. This is a consequence of our method to gather ground truth~(\xref{sec:gt}) that only gives a single matching profile in $SN_2$ for each $a^{1}$. The implications of this limitation on our evaluation is that there might be some matching profiles that we consider as false matches whereas they are not. Since Facebook enforces the policy that users should only have one profile, we believe there are not many such cases and the reliability we measure is likely close to the real-world reliability.

In practice, there are Twitter users that do not have a matching Facebook profile, but our datasets do not contain such cases. To evaluate how matching schemes perform in such scenarios, we test  in~\xref{sec:3step} the reliability of matching schemes when we remove the matching profile from \largescale. 

\section{Generic matching problem}
\label{sec:linker}

%

%

\noindent This section evaluates the reliability of matching schemes based on classifiers aimed at solving the generic case of the matching problem (see~\xref{sec:problem_definition}). %
We build classifiers that are conceptually similar to what previous works have done.  The primary difference between different previous matching schemes is the features and the datasets they used to train and test classifiers, however, they all use traditional classifier such as SVM and Naive Bayes. 
The goal of this section is not to build a matching scheme that is  better than previous ones but to investigate the limits of such schemes in practice.

We first emulate the methodology employed by previous works: we train and test matching schemes with \smallscale, using all attributes.  Since some profile attributes have a high discriminability in the dataset, it is straightforward to build a matching scheme with high reliability. On top of this, there is little difference between the reliability of naive classification techniques and more sophisticated ones. 


We then investigating the reliability of matching schemes in practice by testing them over \largescale.
As expected, the precision of the previously built matching schemes drastically decreases to a point that makes them unusable.  Thereafter, we investigate the reasons behind such poor reliability and we evaluate different strategies to increase the precision and recall in practice. 
The resulting schemes are able to achieve a good precision, but the recall is still low. These results show the inherent difficulty of matching profiles reliably in today's large social networks.

\subsection{Evaluation over \smallscale}
\label{sec:linker_small}

\noindent We use the \smallscale dataset to train and test four classification techniques  to match profiles: Naive Bayes, Decision Trees, Logistic Regression, and SVM. We split \smallscale in two: 70\% for training and 30\% for testing. 

There are two important aspects to handle when training classifiers to match profiles:  ($1$) \emph{classes are very imbalanced} -- there are much more non-matching \profiles than matching \profiles.  Previous works handled this problem by balancing  the training instances by under-sampling the majority class~\cite{He:2009:LID:1591901.1592322}. We also adopt this technique and we randomly sample 850 non-matching profiles from the \smallscale;  ($2$) \emph{features have missing values} -- some attribute values may be unavailable hence the similarity value is missing (e.g., users may choose to omit their location or photo). 
Thus, we must either work with classification techniques that are robust to missing
values (e.g., Naive Bayes) or identify methods to impute the missing values.  

We use 10-fold cross validation on the training data to evaluate the four classifiers with different combinations of parameters and different methods for imputing the missing feature values. We call the four resulting classifiers with the best optimized parameters the \linkernb, \linkersvm, \textsc{Linker-LR} and \textsc{Linker-DT}.


\begin{figure}[t]
\vspace{-2mm}
{\centering
\subfloat[\small{\smallscale}]{\label{fig:prec_rec_random}\includegraphics[width=0.22\textwidth, height=2.7cm]{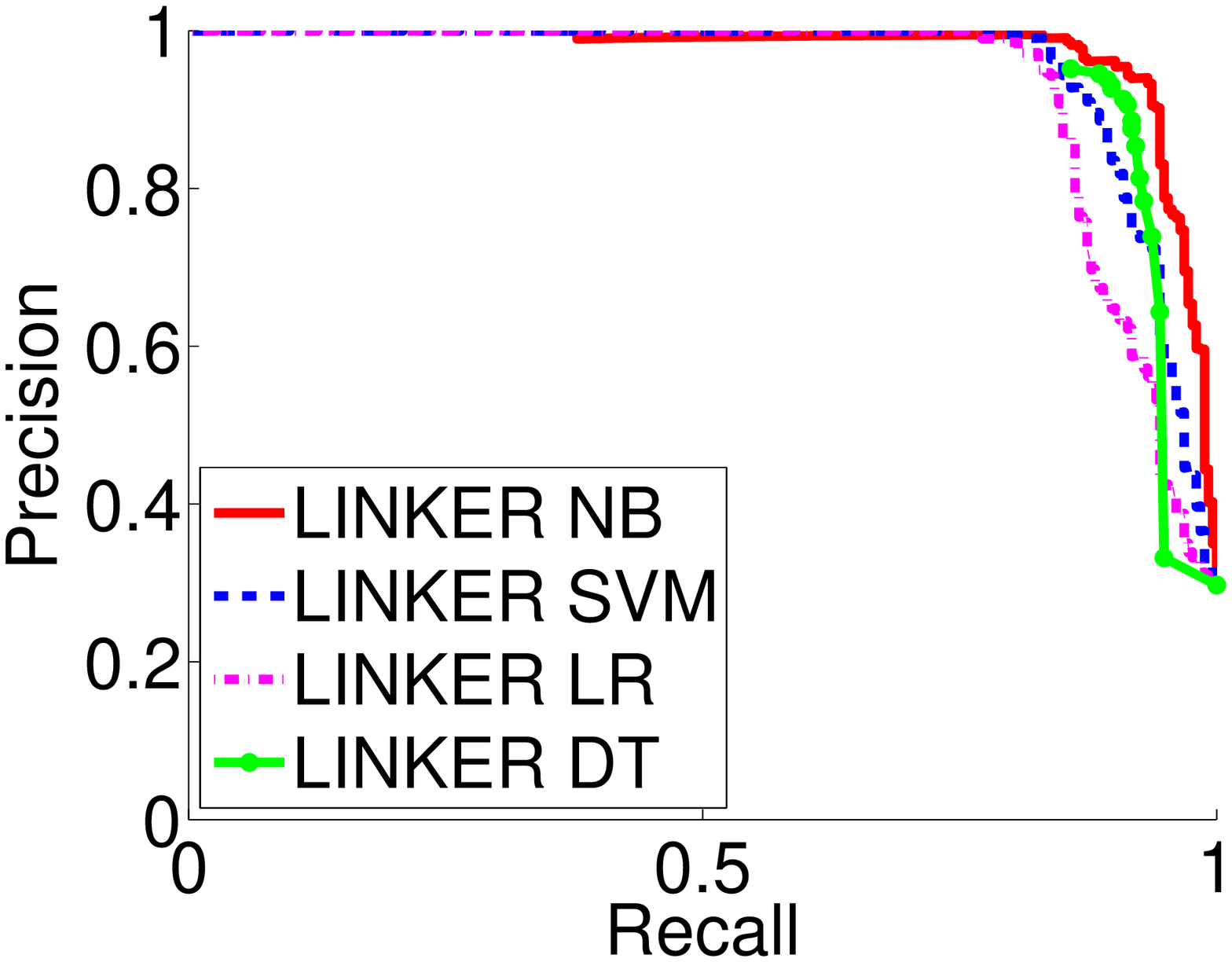}}
\subfloat[\small{\largescale}]{\label{fig:prec_rec_scale}\includegraphics[width=0.22\textwidth, height=2.7cm]{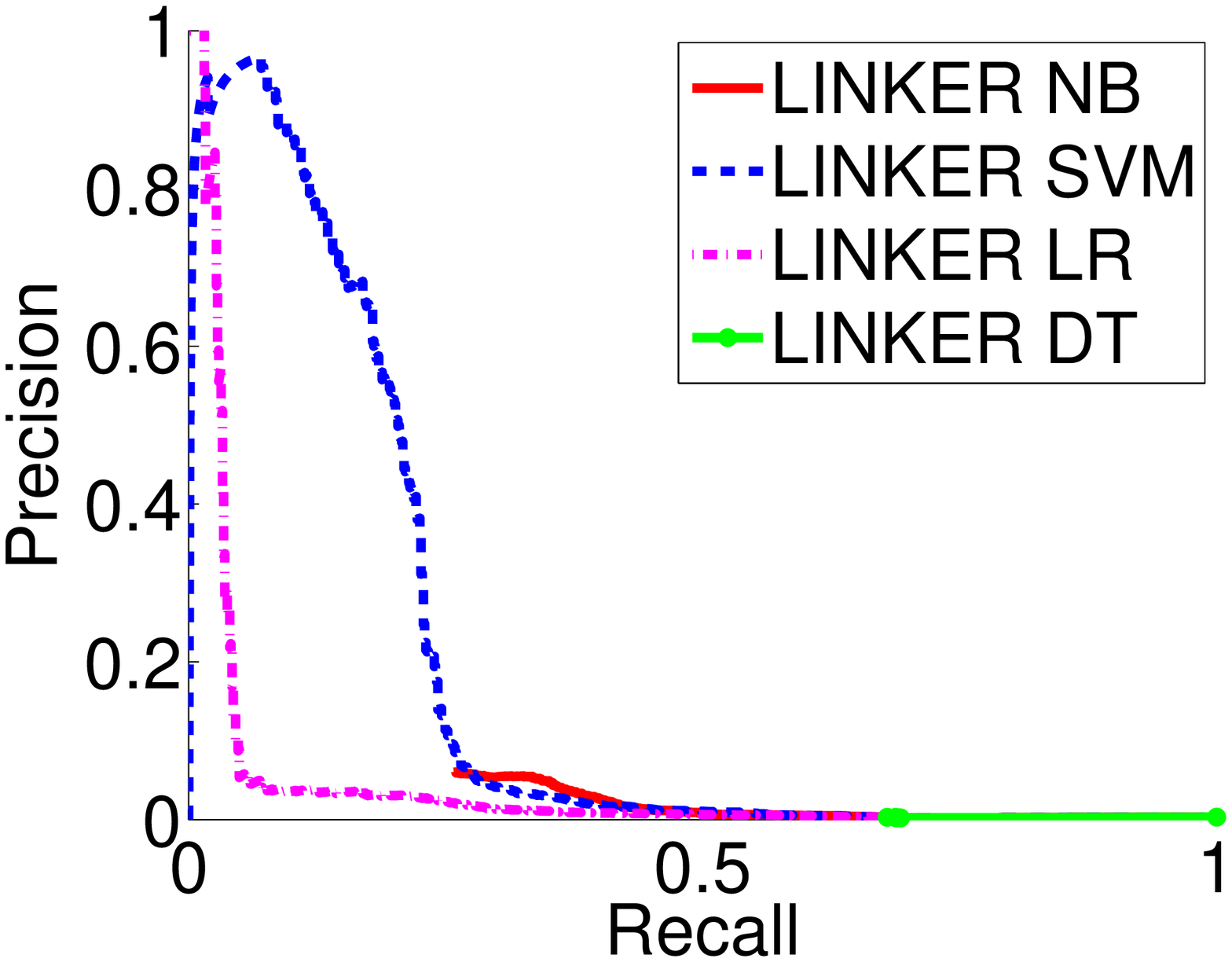}}
\hspace{3mm}

\vspace{-3mm}
\subfloat[\small{\largescale: optimized classifiers}]{\label{fig:prec_rec_scale_retrained}\includegraphics[width=0.22\textwidth, height=2.7cm]{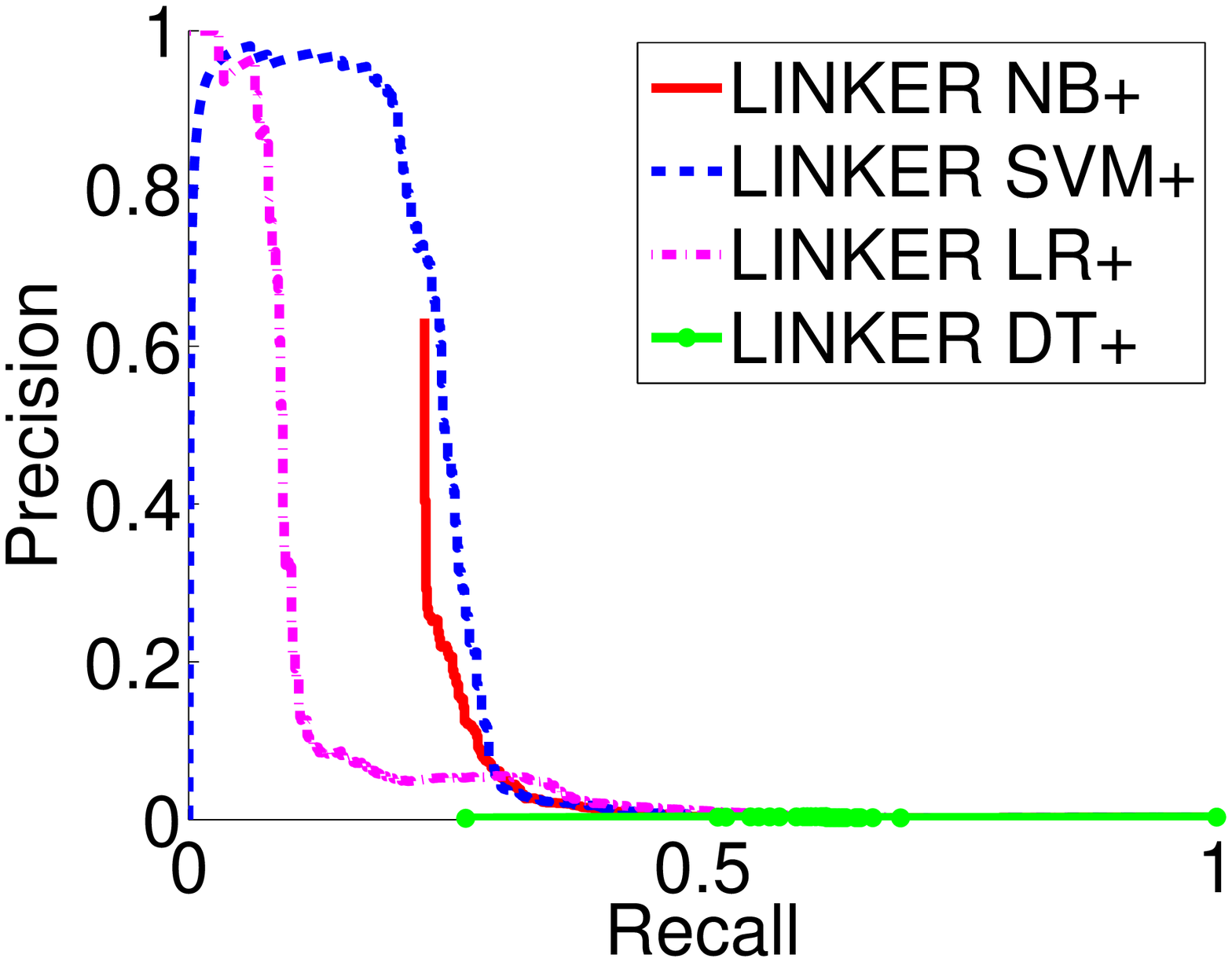}}
\subfloat[\small{\largescale: special case}]{\label{fig:prec_rec_scale_topmatch}\includegraphics[width=0.22\textwidth, height=2.7cm]{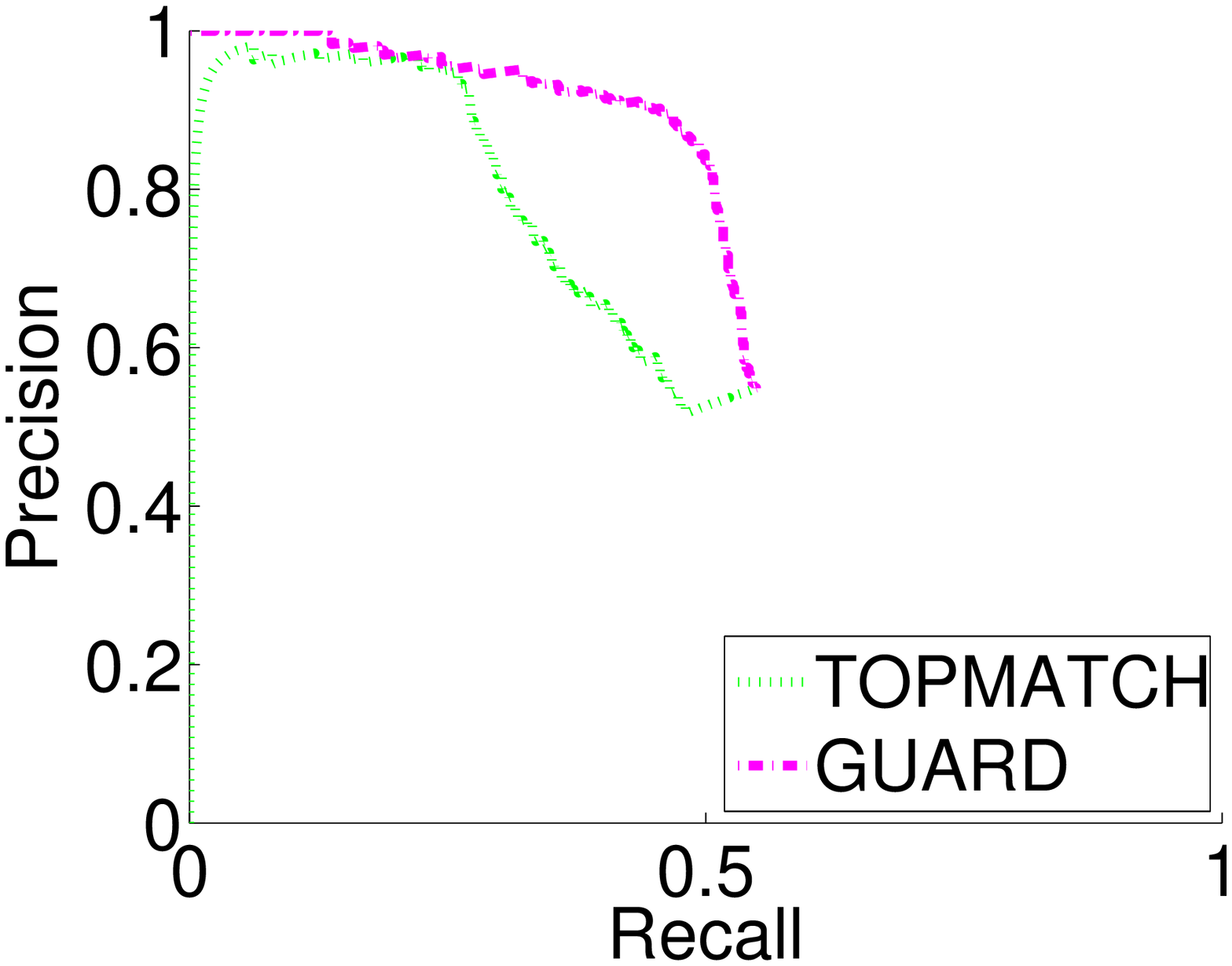}}

\vspace{-2mm}
\caption{\small{Precision and recall tradeoff for matching Twitter to Facebook profiles using different classifiers when evaluated over \smallscale and \largescale.}} 

\label{fig:cdf_prec_rec}

}
\vspace{-0.5\baselineskip}
\end{figure}

  We investigate the  tradeoff between  precision and recall for the different classifiers in Figure~\ref{fig:prec_rec_random}. Our results show that \linkernb out of the box, without imputing the missing values and \linkersvm and \textsc{Linker-DT} when we replace missing values with -1 achieve the highest reliability with a recall over 90\% for a 95\% precision. 
\textsc{Linker-LR} achieves  a lower recall, only 85\% for the same precision. 
Thus, as expected, even out of the box classification techniques such as Naive Bayes are able to achieve a high precision and recall over \smallscale. 


 \vspace{2mm}
 \noindent \minisec{Analysis of matched pairs}
\label{sec:linker_small}
\noindent To understand what pairs of profiles the classifiers are matching, we analyze in Table~\ref{tab:av_cons_small_scale} the availability and consistency of attributes for the \emph{true matches},  the \emph{false matches},  and the \emph{missed matches} (the pairs of matching profiles that are \emph{not} detected by the classifier). 
We use \linkersvm with a threshold on the probability $p$ (outputted by the classifier) corresponding to a 95\% precision (and 90\% recall) to select the true, missed and false matches. 
The table shows that the only matching profiles the \linkersvm is not able to identify are the ones that do not have available and consistent attributes: only 20\% of the missed matches have consistent names and 53\% of missed matches do not have \emph{any} consistent and available attribute (not shown in the table). 
The table also shows that the \linkersvm easily mistakes non-matching profiles form matching profiles if they have either consistent names or friends. While in this dataset this is not problematic, in practice this will lead to many false matches.



\begin{table}[t]
\scriptsize{
\caption{\small{Fraction of true, missed and false matches  that have available and consistent attributes in \smallscale}. 
}
\vspace{-3mm}
{\centering
\begin{tabular}{l|c|c|c|c}
& \multicolumn{4}{c}{Fraction of available and consistent attributes} \\
Feature   & All &  True  & Missed  & False  \\
 &  Matches   &  Matches  & Matches  & Matches  \\
\hline
\hline
Real Name   & 0.77 &  0.91 & 0.20 & 0.62 \\
Screen Name  & 0.38 &  0.46 & 0.07 & 0.09 \\
Location & 0.23 &  0.25 & 0.14 & 0.00 \\
Profile Photo & 0.08 &0.10 & 0.01 & 0.00 \\
Friends & 0.34 & 0.34 & 0.22 & 0.38 \\
\end{tabular}

}
\label{tab:av_cons_small_scale}
}
\vspace{-2\baselineskip}
\end{table}%

\subsection{Evaluation over \largescale}
\label{sec:nary_linker}

%

\noindent Figure~\ref{fig:prec_rec_scale} presents the tradeoff between precision and recall when we evaluate using \largescale the four \sysname classifiers trained on \smallscale.   The figure shows that when matching profiles in practice the reliability of all four classifiers drops significantly compared to \smallscale (presented in Figure~\ref{fig:prec_rec_random}). 
The best classifier on the \smallscale, \linkernb,  achieve only a 4.5\% precision for a 23\% recall when tested on \largescale. The only classifier that achieves a satisfying 95\% precision is \linkersvm, however, the recall is only  15\%.  

These results confirm our intuition that the reliability of a matching scheme over \smallscale fails to capture the reliability of the matching scheme in practice. Worse, the matching scheme that has the best reliability when testing with \smallscale (i.e., the \linkernb) can be amongst the worst in practice.



 \vspace{2mm}
 \noindent \textbf{Optimizing the binary classifiers}

\noindent\minisec{\linkernb{}}
\noindent We investigate the reasons for the low precision of \linkernb in \largescale. The results in Figure~\ref{fig:cdf_new_m_nm}
show that matching \profiles often have consistent names whereas non-matching \profiles (from sample) most often do not; there is no such clear distinction for the other attributes.  Since Naive Bayes assumes that features are independent, the probability that two profiles match will be mainly determined by their name similarity. In a large social network, however, multiple users can have the same name, which will cause \linkernb to output many false matches.


One way to make the classification more accurate is to use two classifiers in cascade instead of one. The first classifier weeds out profiles that are clear non-matches (most of which have different names). Then, the second classifier takes the output of the first and disambiguates the matching profiles out of profiles with similar names. We call this improved classifier \textsc{Linker-NB+}. For more details about this approach please refer to~\cite{Goga}.  

Another approach to make the classification more accurate is to use methods based on joint probabilities such as quadratic discriminant analysis. We prefer to move to SVM which also considers features jointly and is not restricted to quadratic boundaries.

\noindent \minisec{\linkersvm{}}
\noindent \linkersvm has a much higher precision in \largescale than \linkernb. Intuitively, this is because, as 
opposed to Naive Bayes, SVM considers the features jointly and hence can distinguish between pairs of profiles 
with high name similarity that match and pairs of profiles with high name similarity that do not match based on 
other features.  
Nevertheless, previous work has shown that SVM performs suboptimally when using under-sampling to deal with imbalanced datasets~\cite{conf/ecml/AkbaniKJ04}.  By under-sampling the majority class, we are missing informative data points close to the decision boundary. 

To improve the reliability of \linkersvm, we take advantage of the fact that \largescale contains negative examples close to the decision boundary, to enrich our training set. We build a training set that contains \nb positive examples,  \nb negative examples from \smallscale plus \nb negative examples from \largescale. We call the resulting classifier the \hybridsvm. 
Note that if we only use for training negative examples from  \largescale and not from  \smallscale, the resulting classifier will only be able to distinguish the matching profiles out of profiles that look similar and will not be able to distinguish the matching profile out of profiles that are clearly not similar, i.e., it will only work on datasets such as \largescale and not in practice.
For \textsc{Linker-LR}  and \textsc{Linker-DT}  we apply the same retraining technique. 


 \vspace{2mm}
 \noindent \minisec{Evaluation of optimized classifiers}
\noindent Figure~\ref{fig:prec_rec_scale_retrained} shows the tradeoff between precision and recall when using \hybridsvm, \textsc{Linker-NB+}, \textsc{Linker-LR+}, \textsc{Linker-DT+}   on \largescale.  
We can see that \hybridsvm is able to achieve a 19\% recall (4\% improvement over the \linkersvm) for a 95\% precision.\footnote{\small{In \datasetG, \hybridsvm has 50\% recall and 95\% precision.}} 
Also, \textsc{Linker-NB+} achieves a 23\% recall for a 88\% precision, considerably better than \linkernb. Nevertheless, the recall is significantly lower compared with the recall obtained when testing with \smallscale. Thus,  even more sophisticated techniques trained to match profiles in real-word settings fail to match a large fraction of profiles.


 \vspace{2mm}
 \noindent \minisec{Analysis of matched pairs}
\noindent To understand the low recall we obtain in \largescale, we analyze again the availability and consistency of attributes. The precision of \hybridsvm has a sudden drop, to go from a recall of 19\% to 33\%, the precision goes from 95\% to 0.02\%. 
To analyze the drop, we split the pairs of profiles in \largescale in true, missed, and false matches using first a threshold corresponding to a 95\% precision (and a 19\% recall) and then with a threshold corresponding to a 0.02\% precision (and a 33\% recall), see Table~\ref{tab:av_cons_large_scale_95}.  

Contrarily to our expectation, for most attributes but friends, the availability and consistency of  true matches at 0.02\% precision is actually slightly higher than the one at 95\% precision. Only the availability and consistency  of friends  decreases from 91\% at 95\% precision to 68\% at 0.02\% precision. 
This means that, to go from 19\% to 33\% recall we mainly started to match profiles that do not have friends in common. The consequence is that while at 95\% precision, the false matches needed to have friends in common, at 0.02\% precision, false matches no longer need to have  friends in common.  This makes the matching scheme have orders of magnitude more false matches at 33\% recall than at 19\% recall.
Thus, even if the features are highly available and consistent, if they are not discriminable enough, they will allow for many false matches which limits the precision and recall we can achieve in practice.

The results suggest that when matching profiles in practice, to maintain a high precision, we need features that are highly discriminable. Indeed, if we exclude friends (one of the most discriminable attributes) from the features we use for the classification, we can only achieve a 11\% recall for a 90\% precision.

\begin{table}[t]
\scriptsize{
\caption{\small{Fraction of true, missed and false matches  that have available and consistent attributes in \largescale.}} 
\vspace{-3mm}

{\centering
\begin{tabular}{l|c|c|c|c|c|c}
\setlength{\tabcolsep}{1pt}
 & \multicolumn{6}{c}{Fraction of available and consistent attributes} \\

  & \multicolumn{3}{c|}{95\% precision and 19\% recall }  & \multicolumn{3}{c}{0.02\% precision and 33\% recall } \\
Feature  & True  & Missed  & False & True  & Missed  & False   \\
 &  Matches  & Matches  & Matches  &  Matches  & Matches  & Matches  \\
\hline
\hline
Real Name &  0.94 & 0.73 & 0.86 &  0.94 & 0.69 & 1.00 \\
Screen Name &   0.60 & 0.33 & 0.57 &  0.64 & 0.26 & 0.89 \\
Location & 0.32 & 0.21 & 0.14 &  0.41 & 0.15 & 0.01 \\
Profile Photo &    0.14 & 0.07 & 0.57 &  0.16 & 0.04 & 0.07 \\
Friends &  0.91 & 0.17 & 0.57 &  0.68 & 0.14 & 0.00 \\
\end{tabular}

}
\label{tab:av_cons_large_scale_95}
}
\vspace{-2\baselineskip}
\end{table}%
%

\section{Special matching problem}
\label{sec:3step}

\noindent The previous section showed that even fine tuned classifiers are vulnerable to output many false matches in practice. Worse, previous matching schemes are not able to protect against impersonation attacks. In this section, we propose ways to mitigate both of these problems in the special case where we know that there exists \emph{at most one matching profile} in $SN_2$.

\vspace{2mm}
\noindent \minisec{The \psysname}
%
%
\noindent The straw man approach is, for each profile $a^{1}$, to simply return the \profile in $C(a^{1})$ with the highest probability $p$ to be the matching profile given by \hybridsvm, provided that $p$ is larger than a threshold. We call the most similar profile the \psysname. This approach reduces the number of false matches since the matching scheme outputs at most one false match. Figure~\ref{fig:prec_rec_scale_topmatch} displays the tradeoff between precision and recall obtained for different probability thresholds on the $p$ of the \psysname. It shows that \psysname largely improves recall for a given precision: \psysname in \largescale achieves to a 26\% recall for a 95\% precision. 
\vspace{2mm}
\noindent \minisec{The \confidence}
\noindent The strategy of outputting the \psysname considerably increases the recall compared to approaches in~\xref{sec:linker}. However,  it is still vulnerable to output false matches in practice when Twitter users who do not have a Facebook profile. 
Worse, the \psysname is vulnerable to impersonation attacks that also hinder the reliability of the matching scheme. 
We propose next a simple solution that mitigates both of these problems by comparing the probability to be the matching profile of the most similar profile in $C(a^{1})$, $p_{1st}$, and the probability of the second most similar profile, $p_{2nd}$.  
The high level idea is that, to be sure that the most similar profile in $C(a^{1})$ is the matching profile, $p_{1st}$ should be much higher than the probability $p$ of any profile in $C(a^{1})$, i.e., $p_{1st} \gg p_{2nd}$. 

Intuitively, there are two possible scenarios where the \psysname is a false match:
%
%
The first is if an attacker creates an impersonating profile on $SN_2$ that is more similar than the true \matchingprofile. It might be possible to detect these cases as both $p_{1st}$ and $p_{2nd}$ will be high and $(p_{1st} - p_{2nd})$ will be very small.
The second is when the true \matchingprofile $\hat{a}^{2}$ is in $C(a^{1})$
  but a non-matching profile $a^{2} \in C(a^{1})$ is chosen as output
  because the classifier assigns it a higher probability $p$ of being the matching profile (due to the lack of attribute availability and/or consistency).  
  Another case is when $\hat{a}^{2}$ does not exist, forcing the scheme to choose the \nonmatching \profile that is the most  similar to $a^{1}$ as the output. We might detect these cases as $p_{1st}$ and $p_{2nd}$ will not be very high (none of the profiles in $C(a^{1})$ are very similar to $a^{1}$) and $(p_{1st} - p_{2nd})$ will be again very small (none of the profiles in $C(a^{1})$ is much more similar than the rest).

To incorporate the above logic, we design the \confidence which is a binary classifier that takes as input $p_{1st}$ and $p_{2nd}$ and outputs the probability that the \psysname is the \matchingprofile. 
Figure~\ref{fig:prec_rec_scale_topmatch} shows that  the \confidence  increases the recall of the matching scheme to  29\%  for a 95\% precision.  
Although 29\% recall is a big improvement over the recall previously obtained, the recall is still low. This shows that in practice, it is hard to achieve a high recall if we want to have a high precision. 


The matching schemes in~\xref{sec:linker} decide independently for each pair ($a^{1},a^{2}$) where $a^{2} \in C(a^{1})$ whether it is a match or not. In contrast, the strength of the \confidence is that it exploits the structure of $C(a^{1})$ for a given $a^{1}$. In particular, since $C(a^{1})$ depends on $a^{1}$, for a given probability $p$ to be the matching profile of $a^{1}$, the \psysname profile $a^{2}$ will be declared a match for some $a^{1}$ if its attribute values are sufficiently unique, whereas the scheme will return nothing for other $a^{1}$ if the attribute values are too common (e.g., Jennifer Clark that lives in New York). This reduces considerably the false matches and, as we have shown, increases a lot the matching recall for a given precision.

\noindent \minisec{Reliability in the absence of a \matchingprofile}
\noindent To test the reliability of the matching scheme  in the absence of a matching profile, we 
take the \largescale and we remove the matching profiles from the dataset.
Then, we  evaluate the \confidence over the resulting dataset. Ideally, the \confidence should not return any profile as there is no matching profile in the dataset.  Indeed, the \confidence only returns a false match for 1\% of the Twitter \profiles. 
We manually investigate the 1\% cases: in half the returned profile is a false match;  in the other half it is actually a \profile that corresponds to the same person (the returned profiles are either impersonators or people that maintain duplicate \profiles on Facebook).
Thus, the \confidence is reliable when there is no matching \profile in $SN_2$.

\section{Evaluation against humans}
\noindent In this section, we confirm the inherent difficulty to obtain a high recall in matching profiles in practice by comparing our results with results obtained by asking human workers to match profiles.

For this we designed an AMT experiment. We randomly select 200 Twitter-Facebook matching profiles from \datasetFF (that are not used for training the matching schemes).
In each assignment, we give AMT workers a link to a Twitter
\profile as well as links to the 10 most similar Facebook \profiles (we shuffle their position) and we ask AMT workers to choose the matching \profile. We allow  workers to choose that they are unable to identify the \matchingprofile. For each assignment we ask the opinion of three different  workers.  We present the results for majority agreement (two out of three workers decided on the same answer).
We design two versions of the experiment: in the first one if the \matchingprofile is not in $C(a^{1})$,  the \matchingprofile will not be in the list of 10 Facebook \profiles; and a second version, where we always put the matching profile the list of 10 Facebook profiles. 

In the first version of the experiment, AMT workers were able to match 40\% of the Twitter \profiles to their matching profiles and 4\% are matched to the wrong Facebook profile. This means that AMT workers achieve a 40\% recall for a 96\% precision, which is better than the \confidence, but far from a 100\% recall.  In the second version of the experiment, AMT workers were able to match 58\% of Twitter \profiles. Thus, even humans cannot achieve a recall close to 100\% to match profiles in practice.  

\section{Related Works}
\label{sec:existing_schemes}


\noindent We review three primary lines of related research: one
proposing schemes to match user profiles across different social
networks; one focusing on how anonymized user graphs or databases
can be deanonymized to infer user identities; and another about matching entities across databases.




\noindent \minisec{Matching profiles using private user data} 
Balduzzi et al.~\cite{balduzzi09osnProfiling} match
profiles on different social networks using the ``Friend
Finder'' mechanism that social networks provide for users
to find their friends using their email addresses. 
In fact, this is what we use for obtaining our ground truth.
Many sites, however, view Friend Finder as leaking users'
private data and have since limited the number of 
queries a user can make which severely limits the number of profiles one can match.
In contrast, we are interested in understanding the limits of matching profiles by only using public attributes that anyone can access without assuming that we have access to more private data such as the emails of users.

 
\noindent \minisec{Matching profiles using public user data}
A number of previous studies proposed matching schemes that leveraged different attributes of public user data to match profiles, but without systematically understanding their limitations in real-world social networks. 
As a result, previous works overlooked a number of methodological aspects: 
($1$) Most works did not train and test their matching schemes on sampled datasets that preserve the reliability of the original social network. Consequently, the reliability of these schemes drops significantly when evaluated in real-world social networks~\cite{PeritoCKM11,Motoyama,5636108,5272173,conf/socialcom/PeledFRE13,Liu:2013:WNU:2433396.2433457,Zafarani:2013:CUA:2487575.2487648,conf/icwsm/ZafaraniL09}; ($2$) Most works used attributes without analyzing their properties and their  limits to match profiles in practice, consequently, some of these studies use attributes with low availability and thus can only match a small fraction of profiles across a limited number of social networks~\cite{mywww,IofciuFAB11} or use attributes that are prone to give many false matches in practice~\cite{MishariT12}. On the contrary, we propose a framework to analyze attributes and evaluate their potential to match profiles in practice. ($3$) Most studies used biased sets of ground truth users that willingly publish links to their profiles on different social networks. Our analysis reveals that such datasets have attributes that are more available and consistent, consequently, the reliability results of such schemes are overly optimistic~\cite{jain:Iseekfbme:2013:wole,malhotra1,PeritoCKM11,Zafarani:2013:CUA:2487575.2487648}. Other studies assume that all profiles that have the same screen name are matching~\cite{LabitzkeSebastian,IofciuFAB11}. In~\xref{sec:discr} we showed that 20\% of profiles with the same screen name in Twitter and Facebook  are actually not matching.
We further split these studies according to the type of attributes used. 

The closest to our work are a number of schemes that leverage information in the \emph{profiles of users} similar to the attributes we use in this paper~\cite{Motoyama,PeritoCKM11,malhotra1,jain:Iseekfbme:2013:wole,Acquisti,socialsearch,5272173,5636108,northernunsupervised,conf/socialcom/PeledFRE13,Liu:2013:WNU:2433396.2433457,Zafarani:2013:CUA:2487575.2487648,conf/icwsm/ZafaraniL09,socialsearch, LabitzkeSebastian}. 
%
Most schemes work by training classifiers to distinguish between matching and non-matching profiles.  We simulated these approaches in~\xref{sec:linker} and we saw that, because they did not consider the problems that come with matching in practice, the matching schemes are very unreliable when evaluated in real-world social networks. 
 %
A few studies attempted to perform profile matching in practice~\cite{malhotra1,jain:Iseekfbme:2013:wole,Acquisti}. 
These studies, however, just pointed out that profile
matching in practice yields a large number of false matches. 
In contrast, we conduct a systematic analysis of the causes of such false matches and possible ways to eliminate them.

Other schemes use attributes extracted from \textit{user activities} (i.e., the content users generate instead of attributes of the profile)~\cite{mywww,MishariT12,IofciuFAB11}. These schemes reveal how even innocuous
activities of users can help identify a user across social networks. However, these schemes explore attributes with either low availability or low
discriminability, which makes them hard to use in practice without sacrificing reliability.

%
%

\noindent \minisec{De-anonymizing user identities}
\noindent  De-anonymizing user identities and matching user profiles share common methods. 
In fact, our work here is inspired by one of the seminal papers of Sweeney~\cite{sweeneyMedicalDeanon}, which explored the uniqueness
of attributes such as date of birth, postal code, and gender  to de-anonymize medical records of US citizens. 
%
%
Other studies~\cite{narayanan09osn,DBLP:journals/pvldb/KorulaL14} showed the feasibility to
de-anonymize the friendship graph of a social network at large-scale
using the \emph{friendship graph} of another social network as
auxiliary information. The structure of the social graph is certainly  a powerful feature. Nevertheless, in this work, we explicitly assume that we  cannot have access to the entire graph structure of the social networks since we only use public APIs to collect data. We leave as future work how to exploit partial graphs that can be obtained trough APIs to improve matching schemes based on binary classifiers. 



\noindent \minisec{Entity matching} There is a large body of research in the database and information retrieval communities on matching entities across different data sources~\cite{books/daglib/0030287}. 
Conceptually there are many similarities between matching profiles across social networks and matching entities (e.g. the way we compute the similarities between attributes or the adoption of a supervised way to detect matches). However, matching profiles has some specific constraints (e.g., not being able to access all records in $SN_1$) that the entity matching community, to our knowledge,  overlooked.

\section{Conclusion}

\noindent In this paper, we conducted a systematic and detailed
investigation of the reliably of matching user profiles across
real-world online social networks like Twitter and Facebook. Our analysis yielded a number of methodological and measurement contributions.

To understand how profile attributes used by matching schemes affect the overall matching reliability, we proposed a framework that consist of four properties {\it -- Availability, Consistency, Impersonability, and Discriminability (\tname)}. Our analysis showed  that most people maintain the same persona across different social networks -- thus it is possible to match the profiles of many users, however, in practice  there can be a non negligible number of \profiles that belong to different users but have similar attribute values, which leads to false matches.

We showed that the reliability of matching schemes that are trained and tested on reliability non-preserving sampled datasets is not indicative of their reliability in practice.  In fact, traditional matching schemes based on binary classifiers can only achieve a 19\% recall for a 95\% precision  to match Twitter to Facebook profiles in practice.  To avoid these pitfalls we illustrated the right assumptions we can make about the matching problem and the correct methodology to evaluate matching schemes in realistic scenarios.

Finally, we proposed a matching scheme that is able to mitigate impersonation attacks and reduce the number of false matches to achieve a 29\% recall for a 95\% precision. Our matching scheme exploits a special case of the matching problem, namely that there exists at most one matching profile. 
 Although we cannot claim that 29\% is a high recall, humans cannot do much better (they only detect 40\% of matching profiles).

{
\bibliographystyle{abbrv}


\begin{thebibliography}{10}

\bibitem{BingMaps}
{B}ing {M}aps {API}.
\newblock http://www.microsoft.com/maps/developers/web.aspx.

\bibitem{phash}
Phash.
\newblock http://www.phash.org.

\bibitem{spokeo}
Spokeo.
\newblock http://www.spokeo.com/.

\bibitem{spokeolawsuit}
Spokeo lawsuit.
\newblock http://www.ftc.gov/sites/default/files/documents/
  cases/2012/06/120612spokeocmpt.pdf.

\bibitem{Acquisti}
A.~Acquisti, R.~Gross, and F.~Stutzman.
\newblock Faces of facebook: Privacy in the age of augmented reality.
\newblock In {\em BlackHat}, 2011.

\bibitem{conf/ecml/AkbaniKJ04}
R.~Akbani, S.~Kwek, and N.~Japkowicz.
\newblock Applying support vector machines to imbalanced datasets.
\newblock In {\em ECML}, 2004.

\bibitem{AMT_guidelines}
{Get better results with less effort with Mechanical Turk Masters -- The
  Mechanical Turk blog}.
\newblock http://bit.ly/112GmQI.

\bibitem{balduzzi09osnProfiling}
M.~Balduzzi, C.~Platzer, T.~Holz, E.~Kirda, D.~Balzarotti, and C.~Kruegel.
\newblock Abusing social networks for automated user profiling.
\newblock In {\em RAID}, 2010.

\bibitem{books/daglib/0030287}
P.~Christen.
\newblock {\em Data Matching - Concepts and Techniques for Record Linkage,
  Entity Resolution, and Duplicate Detection.}
\newblock Data-centric systems and applications. Springer, 2012.

\bibitem{Cohen03acomparison}
W.~W. Cohen, P.~Ravikumar, and S.~E. Fienberg.
\newblock A comparison of string distance metrics for name-matching tasks.
\newblock In {\em IIWeb}, 2003.

\bibitem{Goga}
O.~Goga.
\newblock {\em Matching User Accounts Across Online Social Networks: Methods
  and Applications}.
\newblock PhD thesis, UPMC, 2014.

\bibitem{mywww}
O.~Goga, H.~Lei, S.~Parthasarathi, G.~Friedland, R.~Sommer, and R.~Teixeira.
\newblock Exploiting innocuous activity for correlating users across sites.
\newblock In {\em WWW}, 2013.

\bibitem{He:2009:LID:1591901.1592322}
H.~He and E.~A. Garcia.
\newblock Learning from imbalanced data.
\newblock {\em IEEE TKDE}, 2009.

\bibitem{IofciuFAB11}
T.~Iofciu, P.~Fankhauser, F.~Abel, and K.~Bischoff.
\newblock Identifying users across social tagging systems.
\newblock In {\em ICWSM}, 2011.

\bibitem{DBLP:journals/pvldb/KorulaL14}
N.~Korula and S.~Lattanzi.
\newblock An efficient reconciliation algorithm for social networks.
\newblock {\em PVLDB}, 2014.

\bibitem{Kreibich:2009:SIL:1855676.1855680}
C.~Kreibich, C.~Kanich, K.~Levchenko, B.~Enright, G.~M. Voelker, V.~Paxson, and
  S.~Savage.
\newblock Spamcraft: An inside look at spam campaign orchestration.
\newblock In {\em LEET}, 2009.

\bibitem{LabitzkeSebastian}
S.~Labitzke, I.~Taranu, and H.~Hartenstein.
\newblock What your friends tell others about you: Low cost linkability of
  social network profiles.
\newblock In {\em SNA-KDD}, 2011.

\bibitem{Liu:2013:WNU:2433396.2433457}
J.~Liu, F.~Zhang, X.~Song, Y.-I. Song, C.-Y. Lin, and H.-W. Hon.
\newblock What's in a name?: An unsupervised approach to link users across
  communities.
\newblock In {\em WSDM}, 2013.

\bibitem{SIFT}
D.~G. Lowe.
\newblock Distinctive image features from scale-invariant keypoints.
\newblock {\em IJCV}, 2004.

\bibitem{malhotra1}
A.~Malhotra, L.~Totti, W.~Meira, P.~Kumaraguru, and V.~Almeida.
\newblock Studying user footprints in different online social networks.
\newblock In {\em CSOSN}, 2012.

\bibitem{MishariT12}
M.~A. Mishari and G.~Tsudik.
\newblock Exploring linkability of user reviews.
\newblock In {\em ESORICS}, 2012.

\bibitem{Motoyama}
M.~Motoyama and G.~Varghese.
\newblock I seek you: searching and matching individuals in social networks.
\newblock In {\em WIDM}, 2009.

\bibitem{narayanan09osn}
A.~Narayanan and V.~Shmatikov.
\newblock De-anonymizing social networks.
\newblock In {\em IEEE S\&P}, 2009.

\bibitem{northernunsupervised}
C.~T. Northern and M.~L. Nelson.
\newblock An unsupervised approach to discovering and disambiguating social
  media profiles.
\newblock In {\em MDSW}, 2011.

\bibitem{jain:Iseekfbme:2013:wole}
P.~K. Paridhi~Jain and A.~Joshi.
\newblock @i seek 'fb.me': Identifying users across multiple online social
  networks.
\newblock In {\em WoLE}, 2013.

\bibitem{peekyou}
Peekyou.
\newblock http://www.peekyou.com/.

\bibitem{conf/socialcom/PeledFRE13}
O.~Peled, M.~Fire, L.~Rokach, and Y.~Elovici.
\newblock Entity matching in online social networks.
\newblock In {\em SocialCom}, 2013.

\bibitem{PeritoCKM11}
D.~Perito, C.~Castelluccia, M.~Ali~K{\^a}afar, and P.~Manils.
\newblock How unique and traceable are usernames?
\newblock In {\em PETS}, 2011.

\bibitem{5636108}
E.~Raad, R.~Chbeir, and A.~Dipanda.
\newblock User profile matching in social networks.
\newblock In {\em NBiS}, 2010.

\bibitem{salesforce}
R.~Schmid.
\newblock Salesforce service cloud -- featuring activision, 2012.
\newblock http://www.youtube.com/watch?v=eT6iHEdnKQ4\&feature=relmfu.

\bibitem{SocialIntelligence}
{Social Intelligence Corp.}
\newblock http://www.socialintel.com/.

\bibitem{sweeneyMedicalDeanon}
L.~Sweeney.
\newblock Weaving technology and policy together to maintain confidentiality.
\newblock {\em Journal of Law, Medicine, and Ethics}, 1997.

\bibitem{5272173}
J.~Vosecky, D.~Hong, and V.~Shen.
\newblock User identification across multiple social networks.
\newblock In {\em NDT}, 2009.

\bibitem{socialsearch}
G.-w. You, S.-w. Hwang, Z.~Nie, and J.-R. Wen.
\newblock Socialsearch: enhancing entity search with social network matching.
\newblock In {\em EDBT/ICDT}, 2011.

\bibitem{conf/icwsm/ZafaraniL09}
R.~Zafarani and H.~Liu.
\newblock Connecting corresponding identities across communities.
\newblock In {\em ICWSM}, 2009.

\bibitem{Zafarani:2013:CUA:2487575.2487648}
R.~Zafarani and H.~Liu.
\newblock Connecting users across social media sites: A behavioral-modeling
  approach.
\newblock In {\em KDD}, 2013.

\end{thebibliography}
}

\appendix

\section{Proofs from Section~3}
\label{ap:proof}
\subsection{Effective discriminability formula}

\noindent In this section, we justify that, if we assume that the impersonating profiles are independent from the other non-matching profiles, we have $\tilde{D} = D \cdot (p_{nI} + nI \cdot p_{I})$. We first apply the complete probability formula: 
\begin{align}
\nonumber \tilde{D} & \! = \! Pr\Big(\max_{a^{2}: \nonmatch} s(v^{1},v^{2}) <  th\Big) \\ 
\nonumber & \! = \! Pr\Big(\max_{a^{2}: \nonmatch} s(v^{1},v^{2}) <  th  \big| a^{1} \textrm{ not impersonated}\Big) p_{nI} \\ 
\label{eq1} & \! + \! Pr\Big( \! \max_{a^{2}: \nonmatch} \!\! s(v^{1},v^{2}) \! < \! th  \big| a^{1} \textrm{ impersonated}\Big) p_{I}.
\end{align}
Then, we observe that the $\max$ on all non-matching profile is smaller than $th$ iif both the $\max$ on all non-matching profile that are not impersonating $a^{1}$ and the max on the impersonators of $a^{1}$ are smaller than $th$. That is:
\begin{align}
\nonumber & Pr\Big( \max_{a^{2}: \nonmatch}  s(v^{1},v^{2})  < th  \big| a^{1} \textrm{ impersonated}\Big) \\ 
\label{eq2} & = Pr\Big( \max_{a^{2}: \nonmatch \atop \nonimpers } s(v^{1},v^{2}) < th, \\
\nonumber & \;\;\;\;\;\;\;\; \max_{a^{2}: \impers} s(v^{1},v^{2}) < th \Big|  a^{1} \textrm{ impersonated}\Big). 
\end{align}
If the impersonating profiles are independent from the other non-matching profiles, then the joint probability equals the product of probabilities: 
\begin{align*}
& Pr\Big( \max_{a^{2}: \nonmatch}  s(v^{1},v^{2})  < th  \big| a^{1} \textrm{ impersonated}\Big) \\ 
& = Pr\Big( \max_{a^{2}: \nonmatch \atop \nonimpers } s(v^{1},v^{2}) < th \Big|  a^{1} \textrm{ impersonated}\Big) \cdot \\
& \;\;\;\;\;  Pr\Big( \max_{a^{2}: \impers} s(v^{1},v^{2}) < th \big|  a^{1} \textrm{ impersonated}\Big) \\ 
& = D\cdot nI;
\end{align*}
which, given \eqref{eq1}, shows that $\tilde{D} = D \cdot (p_{nI} + nI \cdot p_{I})$.

\subsection{Proofs of Theorem~1}

\noindent In this section, we give a proof of Theorem~\ref{theorem}. Recall that $th$ is a threshold parameter in $[0, 1]$ such that the classifier declares a match between $a^1$ and $a^2$ if $s(v^{1}, v^{2}) > th$. 

\bigskip

To show \textit{(i)}, first recall the definition of recall: 
\begin{equation*}
recall = Pr \left( s(v^{1}, v^{2}) > th | \match  \right). 
\end{equation*}
Then, we have 
\begin{align*}
recall = & Pr \left( s(v^{1}, v^{2}) > th | \match, v^{1} \textrm{ and } v^{2} \textrm{ available}  \right)\\
 & \cdot Pr (v^{1} \textrm{ and } v^{2} \textrm{ available} | \match)\\
 & +  Pr \left( s(v^{1}, v^{2}) > th | \match, v^{1} \textrm{ or } v^{2} \textrm{ not available}  \right)\\
 & \cdot Pr (v^{1} \textrm{ or } v^{2} \textrm{ not available} | \match).
\end{align*}
By convention, $s(v^{1}, v^{2}) = 0$ if $v^{1} \textrm{ or } v^{2} \textrm{ not available}$ (a pair is never declared a match by the classifier if either value is missing). Therefore, $Pr \left( s(v^{1}, v^{2}) > th | \match, v^{1} \textrm{ or } v^{2} \textrm{ not available}  \right) = 0$ and we have $recall = C\cdot A$ by definition of $C$ and $A$.

\bigskip

To show \textit{(ii)}, first recall the definition of precision:  
\begin{equation*}
precision = Pr \left( \match | s(v^{1}, v^{2}) >th \right). 
\end{equation*}
To ease the equations reading, we simplify the notation of $\match$ into simply $\matchs$ and similarly for $\nonmatch$. By application of Bayes formula, we compute
\begin{align*}
& precision \\
& =  \frac{Pr (\matchs, s(v^{1}, v^{2}) >th) }{Pr (\matchs, s(v^{1}, v^{2}) >th) + Pr (\nonmatchs, s(v^{1}, v^{2}) >th)}\\
& = \frac{recall \cdot Pr (\matchs)}{recall \cdot Pr (\matchs) + Pr (\nonmatchs, s(v^{1}, v^{2}) >th)}. 
\end{align*}
Let $n_2$ denote the number of profiles in $SN_2$\footnote{Note that $n_2$ includes impersonating profiles and hence formally is a random variable. Rigorously, we should condition on the value of $n_2$ and then take the expectation; however the result would be unchanged hence we omit this detail for a lighter presentation.}. By the assumption of Theorem~\ref{theorem}-\textit{(ii)}, we have $Pr(\matchs) \le 1/n_2$, so that, since $Pr (\nonmatchs, s(v^{1}, v^{2}) >th)\ge 0$, we get
\begin{align*}
precision \le \frac{recall}{recall + n_2 \cdot Pr (\nonmatchs, s(v^{1}, v^{2}) >th)}. 
\end{align*}
Moreover, by definition of $\tilde{D}$, we have 
\begin{equation*}
Pr (\nonmatchs, s(v^{1}, v^{2}) >th) \ge \frac{1-\tilde{D}}{n_2},
\end{equation*}
which gives 
\begin{align*}
precision \le \frac{recall}{recall + 1-\tilde{D}}
\end{align*}
and concludes the proof of Theorem~\ref{theorem}-\textit{(ii)}.

\bigskip

We now show \textit{(iii)}, by making three observations: 
\begin{trivlist}
\item[a.] First, observe that, from \textit{(i)}, we directly get that $recall=1$ iif $A = C = 1$. 
\item[b.] Second, observe that $precision =1$ iif $\tilde{D}=1$. Indeed, we have $precision = 1$ iif $Pr(\nonmatch, s(v^{1}, s^{2})>th) = 0$, which is equivalent to $Pr(\max_{a^{2}: \nonmatch} s(v^{1},v^{2}) >  th) = 0$ and hence to $\tilde{D}=1$. 
\item[c.] Third, observe that $D = nI = 1$ implies $\tilde{D}=1$ and that, if $p_{I}>0$, the converse holds too. We show the two separately. \\
$(\Rightarrow):$ Assume that $D = nI = 1$. The result follows from the following facts: if $D=1$ then the first term of \eqref{eq1} multiplying $p_{nI}$ is $1$; and from \eqref{eq2}, if $D=1$ and $D=1$ and $nI=1$, then the second term of \eqref{eq1} multiplying $p_{I}$ is $1$. Therefore, $\tilde{D}=1$. \\
$(\Leftarrow):$ Assume that $\tilde{D}=1$ and $p_I>0$. If $D<1$, then both terms of \eqref{eq1} multiplying $p_{nI}$ and $p_I$ are strictly smaller than one which contradicts $\tilde{D}=1$. Therefore $D=1$. If $nI<1$, the second term of \eqref{eq1} multiplying $p_{I}$ is strictly smaller than one which contradicts $\tilde{D}=1$ since $p_I>0$. Therefore $nI=1$.  \\
\end{trivlist}
The combination of these three observations implies Theorem~\ref{theorem}-\textit{(iii)}. (In fact, these three observations give more detailed results on the impact of ACID on precision and recall than what is summarized in Theorem~\ref{theorem}-\textit{(iii)}.)


\section{Attribute similarity metrics}
\label{ap:sim}
\vspace{2mm}
\noindent \minisec{Name similarity} Previous work in the record
linkage community showed that the \emph{Jaro string distance} is the
most suitable metric to compare similarity between names both in the
offline and online worlds~\cite{Cohen03acomparison, PeritoCKM11}. So
we use the Jaro distance to measure the similarity between real names
and screen names.

\vspace{2mm}
\noindent \minisec{Photo similarity} Estimating photo similarity is
tricky as the same  photo can come in different formats.  
To measure the similarity of two photos
while accounting for image transformations, we use two matching
techniques: ($i$) \emph{perceptual hashing}, a technique originally
invented for identifying illegal copies of copyrighted content that
works by reducing the image to a transformation-resilient
``fingerprint'' containing its salient characteristics ~\cite{phash}
and ($ii$) {\it SIFT}, a size invariant algorithm that detects local
features in an image and checks if two images are similar by counting
the number of local features that match between two
images~\cite{SIFT}.
%
We use two different algorithms for robustness. The perceptual
hashing technique does not cope well with some images that are
resized, while the SIFT algorithm does not cope well with computer
generated images. 

\vspace{2mm}
\noindent \minisec{Location similarity} For all profiles, we have the
textual representations of the location, like the name of a
city. Since social networks use different formats for this
information,  a simple textual comparison will be inaccurate. Instead, we
convert the location to latitude/longitude coordinates by submitting
them to the Bing API~\cite{BingMaps}.  We then compute the similarity between two locations as the actual geodesic distance between
the corresponding coordinates.

\vspace{2mm}
\noindent \minisec{Friends similarity} The similarity score is the
number of common friends between two profiles. We consider that two
profiles have a common friend if there is a profile with the same
screen name or real name in both friend lists. A more complex but potentially more
accurate method would have been to apply a matching scheme for each friend recursively 
taking other features beside screen name and real name into account. 
As we will see, however, given two small lists of profiles on different social networks, real names and screen 
names alone can accurately identify matching \profiles. Complementary, we could divide the number of common friends by the total number of friends. Preliminary results showed no particular improvement in doing so.

\end{document}